\documentclass[journal,10pt,twocolumn]{IEEEtran}
\IEEEoverridecommandlockouts
\usepackage[sort]{cite}
\ifCLASSINFOpdf
\usepackage[pdftex]{graphicx}
  \graphicspath{{graphics/pdf/}{graphics/png}}
\else
\fi
\usepackage[table,xcdraw]{xcolor}
\usepackage{epstopdf}
\usepackage[cmex10]{amsmath}
\usepackage{cancel} 
\usepackage{array}
\usepackage{mdwmath}
\usepackage{mdwtab}
\usepackage{eqparbox}
\usepackage{stfloats}
\usepackage{url}
\usepackage[T1]{fontenc}
\usepackage{amsfonts}
\usepackage{rotating}
\usepackage{amsthm}
\usepackage[hidelinks]{hyperref}
\usepackage{epstopdf}

\usepackage{pgfplots}
\pgfplotsset{compat=newest} 
\pgfplotsset{plot coordinates/math parser=false}
\usepackage{float}
\usepackage{algorithm}
\usepackage{algpseudocode}
\usepackage{tikz}
\usepackage{multirow}
\usepackage{ctable}
\usepackage{exscale}
\usepackage{grffile}
\usepackage{amssymb}
\usepackage{soul}
\usepackage{grffile}
\usepackage{todonotes}
\usepackage{subcaption}
\usetikzlibrary{plotmarks}
\usetikzlibrary{arrows.meta}
\usepgfplotslibrary{patchplots}
\begin{document}
\title{On the Achievable Rate of MIMO Narrowband PLC with Spatio-Temporal Correlated Noise} 
\author{Mohammadreza~Bakhshizadeh~Mohajer, Sadaf~Moaveninejad, Atul~Kumar, Mahmoud~Elgenedy, Naofal~Al-Dhahir,~\IEEEmembership{Fellow,~IEEE},
Luca~Barletta,~\IEEEmembership{Member,~IEEE}, and~Maurizio~Magarini,~\IEEEmembership{Member,~IEEE}

\thanks{M. B. Mohajer, L. Barletta and M. Magarini are with the Department of Electronics, Information and Bioengineering, Politecnico di Milano, I-20133, Milan, Italy e-mail: mohammadreza.bakhshizadeh@mail.polimi.it,
luca.barletta@polimi.it,
maurizio.magarini@polimi.it. 

S. Moaveninejad is with the Department of Neuroscience, Università degli Studi di Padova, Padova, Italy. 

A. Kumar is with Department of Electronics Engineering, IIT (BHU)
Varanasi, Varanasi 221005, India.

M. Elgenedy is with Qualcomm, Worcester, MA, USA.

N. Al-Dhahir is with Department  of Electrical  and  Computer  Engineering,  The  University  of  Texas  at  Dallas, Richardson,  TX  75080  USA.
}

}

\maketitle

\begin{abstract}
Narrowband power line communication (NB-PLC) systems are an attractive solution for supporting current and future smart grids. A technology proposed to enhance data rate in NB-PLC is multiple-input multiple-output (MIMO) transmission over multiple power line phases.
To achieve reliable communication over MIMO NB-PLC, a key challenge is to take into account and mitigate the effects of temporally and spatially correlated 
cyclostationary noise. Noise samples in a cycle can be divided into three classes with different distributions, \emph{i.e.} Gaussian, moderate impulsive, and strong impulsive. 
However, in this paper we first show that the impulsive classes in their turn can be divided into sub-classes with normal distributions and, after deriving the theoretical capacity, two noise sample sets with such characteristics are used to evaluate achievable information rates: one sample set is the measured noise in laboratory and the other is produced through MIMO frequency-shift (FRESH) filtering. 
The achievable information rates are attained by means of a spatio-temporal whitening of the portions of the cyclostationary correlated noise samples that belong to the Gaussian sub-classes. The proposed approach can be useful to design the optimal receiver in terms of bit allocation using waterfilling algorithm and to adapt modulation order. 

\end{abstract}


\section{Introduction}

Narrowband power line communication (NB-PLC) has been receiving increased popularity~\cite{interest_in_plc}, as a strong candidate to connect meters with local utilities~\cite{metering,Elgenedy-2015-mitigation,smartgrid_plc_review}, forming essential parts of smart grid (SG), which will provide information regarding power consumption and demand. The intelligent monitoring and control of the energy flows in the electric network represents the first path for transition from conventional power networks towards SG~\cite{Noise_survey,SmartMetering2022}. 

The operating frequency of NB-PLC systems is in the $3-500\,$kHz range. In Europe, the allocated bandwidth for NB-PLC is determined by the Comité Européen de Normalisation Électrotechnique (CENELEC) signaling standard. This standard classifies frequencies from $3\,$kHz up to $148.5\,$kHz into four sub-bands. CENELEC-A band, which encompasses frequencies in the $3-95\,$kHz range, is assigned to electricity suppliers and distributors~\cite{pittolo2014performance}. The other three bands, namely B, C, and D, are in the $95-148.5\,$kHz range and are reserved for energy customers. Communication with narrowband frequencies over power line was initially proposed for single-carrier modulation and single-input single-output (SISO) transmission, which provides low data rates on the order of few kilobits per second (kbps). Higher data rates (HDR) were obtained by introducing orthogonal frequency-division multiplexing (OFDM)~\cite{Noise_survey}. An HDR is required to improve the efficiency for new SG applications, such as tele-management services, and to decrease the time to access a meter's data~\cite{moaveninejad2017ber}. To achieve further enhancements of the data rate in NB-PLC, multiple-input multiple-output (MIMO) transmission is considered as a promising technology. In many regions of the world there is a three-wire installation for electric power delivery that allows realization of MIMO transmission in PLC, where couples of wires are mapped on multiple feeding and receiving ports. The introduction of MIMO allows the limited bandwidth of NB-PLC to be exploited more efficiently \cite{MIMO_advantage,schwager2011mimo}.

In addition to HDR, establishing communication services in SG also requires highly reliable transmission. To achieve high reliability, one of the fundamental issues that must be overcome is the distortion introduced by the time-varying and correlated noise \cite{nassar2012cyclostationary}. The main contribution of this paper is to evaluate the channel capacity of a MIMO NB-PLC by taking into account the correlation among noise samples both in the temporal and spatial domains. Noise in NB-PLC, which can be viewed as the superposition of different sources~\cite{Noise_survey}, has periodic temporal and spectral properties that are synchronous to AC mains cycle \cite{nassar2012cyclostationary,Nieman-cyclic-spectral-2013}. In NB-PLC transmission, the symbol duration, and consequently the packet length, tends to be long because of the relatively narrow bandwidth, and thus the periodic features of the noise cannot be ignored \cite{katayama2006mathematical}. When moving to MIMO, another challenge is represented by the spatial correlation of the noises affecting multiple phases~\cite{elgenedy2016frequency}. A well established model that considers the effect of correlated cyclostationary noise in NB-PLC is based on the use of frequency-shift (FRESH) filtering. FRESH filtering was used in~\cite{elgenedy2016cyclostationary} for SISO NB-PLC to reproduce temporally correlated noise samples, and then was extended in~\cite{elgenedy2016frequency} to MIMO NB-PLC with both temporal and spatial correlation between noise samples. The FRESH model was validated with real measurements obtained in lab experiments, which are also used in this paper to estimate the capacity. In ~\cite{moaveninejad2019gaussian}, it was shown that the noise samples generated by the FRESH filter have an amplitude distribution in each cycle that can be classified into three different classes. To the best of the authors' knowledge, \cite{moaveninejad2019gaussian} was the first work that introduced a classification of the different temporal portions of the cyclostationary noise in MIMO NB-PLC according to the measured variance, which is more accurate than the temporal region classification based on visual inspection proposed in~\cite{Elgenedy-2015-mitigation}, and~\cite{nassar2012cyclostationary}. The analysis done in \cite{moaveninejad2019gaussian} revealed that the probability density function (pdf) of one class fits well with that of a Gaussian distribution while the other two classes are non-Gaussian and can be considered as strong and moderate impulsive. Note that, other classifications with more parameters could be applied by using machine learning approaches~\cite{tonello2019machine,righini2019automatic}. Moreover by applying machine learning techniques  \textit{i.e.} neural networks, it is possible to predict the behavior of the noise in the next cycles.
The predicted samples are used to update the classification, characterization, and spatio-temporal whitening operations~\cite{tonello2019machine,MLNoiseReproduction}.

In this paper, to evaluate the capacity, a Gaussianity test is used to divide each cycle of the noise into smaller portions with normal distribution. Then, the Shannon capacity formula is used to compute the achievable information rate. This is the first time that both spatial and temporal aspects of noise correlation are considered at the same time to evaluate the capacity in MIMO NB-PLC. In~\cite{pittolo2014performance} and \cite{rende2011noise} only the spatial correlation was considered to evaluate the capacity for MIMO broadband PLC. The authors of~\cite{nikfar2014mimo} analyzed the effect of the spatial correlation on the capacity of a MIMO NB-PLC system in the presence of Middleton class-A noise. The same noise model was utilized in~\cite{ergodicOFDM} to derive the ergodic capacity of an OFDM NB-PLC system. However, the temporal correlation between samples was ignored in the Middleton noise model. The temporal correlation was considered in~\cite{shlezinger2015capacity} to evaluate the capacity of a SISO NB-PLC that accounts for periodic properties of both channel and noise. In this regard, a linear periodically time varying (LPTV) channel model was assumed in the presence of additive cyclostationary Gaussian noise (ACGN). Numerical results were shown for the capacity of both a flat channel and LPTV channel in the presence of noise samples generated by models proposed by Katayama \textit{et al.} in~\cite{katayama2006mathematical} and the one given in the IEEE 1901.2 standard~\cite{standard}. The authors in \cite{STBCbasedPerformance} have studied the performance of space-time block codes (STBC) in receive diversity of MIMO NB-PLC systems considering Rayleigh fading channel and cyclostationary noise generated using the model in~\cite{katayama2006mathematical}. The study assumes that the noise in each channel is independent and disregards spatial correlation. Recently, \cite{SamplingRateAndCapacity} has shown how the sampling rate and sampling time offset can affect the capacity of a memoryless channel in the presence of cyclostationary Gaussian noise. 

The rest of this paper is organized as follows. Section~\ref{sec:background} provides a review on FRESH filtering in both SISO and MIMO systems. Then, the theoretical channel capacity for the MIMO NB-PLC in the presence of cyclostationary spatio-temporal correlated noise is derived in Sec.~\ref{sec:cap_corr}: This section provides a method for spatial and temporal whitening of the correlated noise. In Sec.~\ref{sec:numerical_evaluation}, a lower bound on the capacity is evaluated numerically for two sample sets, one measured in the laboratory and the other one generated using the FRESH filtering model defined in~\cite{elgenedy2016frequency}, considering the Gaussian noise approximation. Section~\ref{sec:numerical_evaluation} also presents a description of the system model, and a classification of the noise together with the test of Gaussianity to define the length of portions of noise samples with normal distribution. 
Finally, the paper is concluded in Sec.~\ref{sec:Conclusion}.

\section{Background on Cyclostationary Noise Models for SISO and MIMO NB-PLC}
\label{sec:background}

Katayama \textit{et al.} \cite{katayama2006mathematical} proposed a cyclostationary model for the noise in SISO NB-PLC systems, where the samples are described by means of zero-mean Gaussian random variables with time-varying variances. Later, another cyclostationary model was introduced by Nassar \textit{et al.}~\cite{nassar2012cyclostationary} that, in addition to a time-domain characterization, also gives a frequency-domain characterization. 
Authors of~\cite{nassar2012cyclostationary} divided the NB-PLC noise into $M$ temporal stationary regions, where temporal correlation of the noise samples is limited to each region. The model does not take into account the correlation among the NB-PLC noise samples in different regions. This correlation, which is ignored in the standard, was first observed in~\cite{elgenedy2016cyclostationary}.
Moreover, considering MIMO NB-PLC systems, a noise model needs also to include the effect of correlation between different phases. Yet, to the best knowledge of the authors, \cite{elgenedy2016frequency} is the only work that gives a model where the cyclostationary noise of a MIMO NB-PLC is correlated both in time and space, which is called a MIMO FRESH filter. In the following, we will describe its genesis by starting from the SISO case.
\subsection{FRESH Filtering for SISO}
\begin{figure}
	\centering
	\includegraphics[width=.35\textwidth]{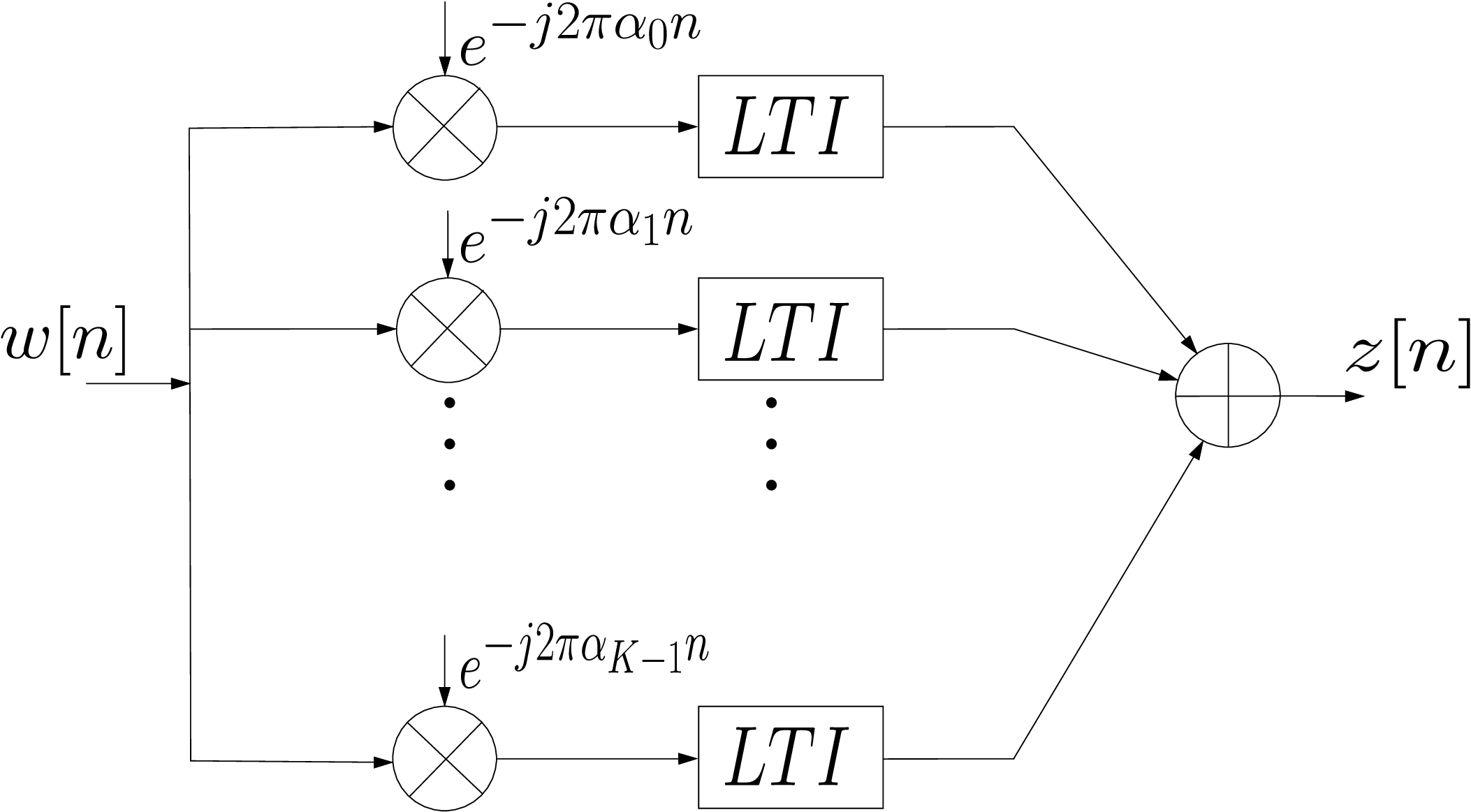}
	\caption{SISO FRESH filtering Model.}
	\label{fig:noise-fresh-siso}
 
\end{figure}
FRESH filtering is a technique that was originally proposed by Gardner in~\cite{wiener-1993-sisofresh} to implement linear time-varying SISO filters. It can be considered as the cyclostationary counterpart of linear time invariant (LTI) filtering. With this in mind, the authors of~\cite{elgenedy2016cyclostationary} proposed a cyclostationary noise model for SISO systems based on FRESH filtering.

The generation of the noise can be described by starting from the input-output relationship of an LPTV filter, given by
\begin{equation}\label{eq:LPTV}
    z[n] = \sum\limits_{l= - \infty}^{+\infty} g[n,l]w[l],
\end{equation}
where $z[n]$ are the generated samples of the cyclostationary noise, $w[l]$ is an additive white Gaussian noise (AWGN) process that serves as input excitation, and $g[n,l]$ is the impulse response of the filter which is defined as
\begin{equation}\label{eq:shifted_LPTV}
    g[n,l] = \sum\limits_{k= 0}^{K-1} {\tilde g_{k}[n-l]} e^{-j2\pi \alpha_{k} l}.
\end{equation}
In \eqref{eq:shifted_LPTV}, ${\tilde g_{k}[n-l]}$ is the $k$th Fourier series coefficient of $g[n,l]$, $\alpha_{k} = \frac{k}{K}$, $k = 0, 1, \cdots ,K-1$ is the $k$th cyclic frequency, and $K$ is the period of the cyclostationary noise which is also the number of branches of the FRESH filter. After substitution, \eqref{eq:LPTV} can be written as
\begin{equation}
    z[n] = \sum\limits_{k = 0}^{K-1} \sum\limits_{l = -\infty}^{+\infty} {\tilde g_{k}[n-l]} w_{k}[l],
\end{equation}
where $w_{k}[l] = w[l] e^{-j 2 \pi \alpha_{k} l}$. Hence, the system performs LTI filtering on frequency shifted versions of the input signal $w[l]$, which is then modeled as a bank of LTI filters. The corresponding structure is shown in Fig.~\ref{fig:noise-fresh-siso}.

\subsection{FRESH Filtering for MIMO}
\label{subsubsec:FRESH-MIMO}
In \cite{elgenedy2016frequency}, the SISO FRESH filter architecture was generalized to the MIMO case.
In contrast to SISO, the MIMO system involves multiple input signals for each phase, which are shifted in $K$ branches of the FRESH filter and subsequently filtered by an LTI filter.

\begin{figure}
	\centering
	\includegraphics[width=\columnwidth]{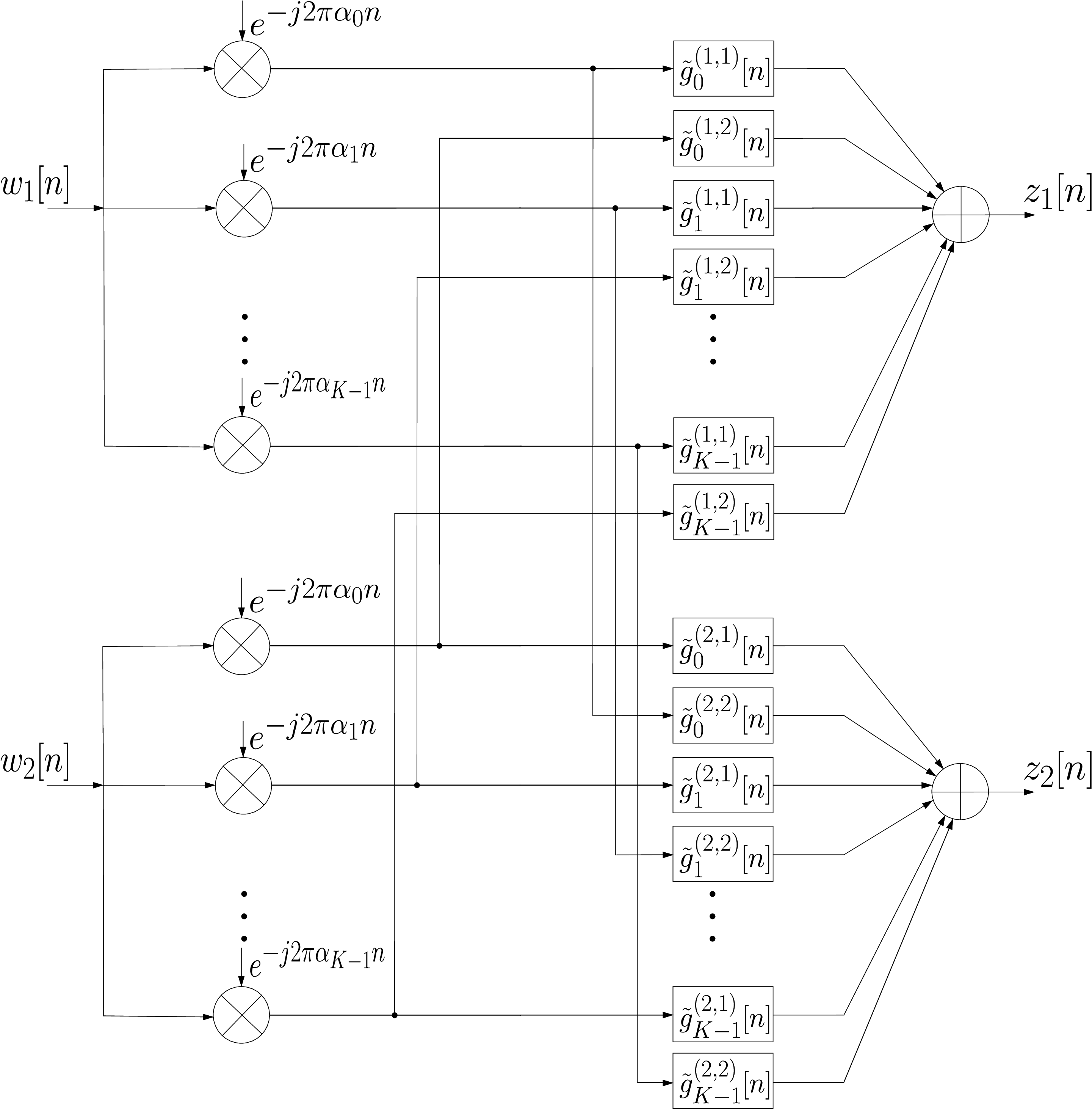}
           \caption{FRESH filter for cyclostationary noise generation in a $2\times 2$ MIMO system.}
	\label{fig:noise-fresh-mimo}
    
\end{figure}

Figure~\ref{fig:noise-fresh-mimo} shows the structure of the FRESH filtering for a $2\times2$ MIMO system, where for the $k$-th branch there are four LTI filters, namely $ {{\tilde g}^{(1,1)}_{k}}[n], {{\tilde g}^{(1,2)}_{k}}[n], {{\tilde g}^{(2,1)}_{k}}[n], {{\tilde g}^{(2,2)}_{k}}[n]$, which are arranged in the matrix
\begin{equation}
{\tilde g_{k}}[n]= 
\left[ {\begin{array}{*{20}c}
    {{\tilde g}^{(1,1)}_{k}}[n] & {{\tilde g}^{(1,2)}_{k}}[n]  \\
     {{\tilde g}^{(2,1)}_{k}}[n] &   {{\tilde g}^{(2,2)}_{k}}[n] \\    
 \end{array} } \right].
\end{equation}
By summing the results related to each stream, the two outputs $z^{(1)}[n]$ and $z^{(2)}[n]$ that reproduce the cyclostationary noises are formed. Authors of~\cite{elgenedy2016frequency} validated the accuracy of their model by using a normalized mean squared error (NMSE) metric between the cyclic auto/cross correlation of the generated and measured noise samples.

\section{Channel Capacity with Correlated Noise}
\label{sec:cap_corr}

\begin{figure}
	\centering
	\includegraphics[width=.35\textwidth]{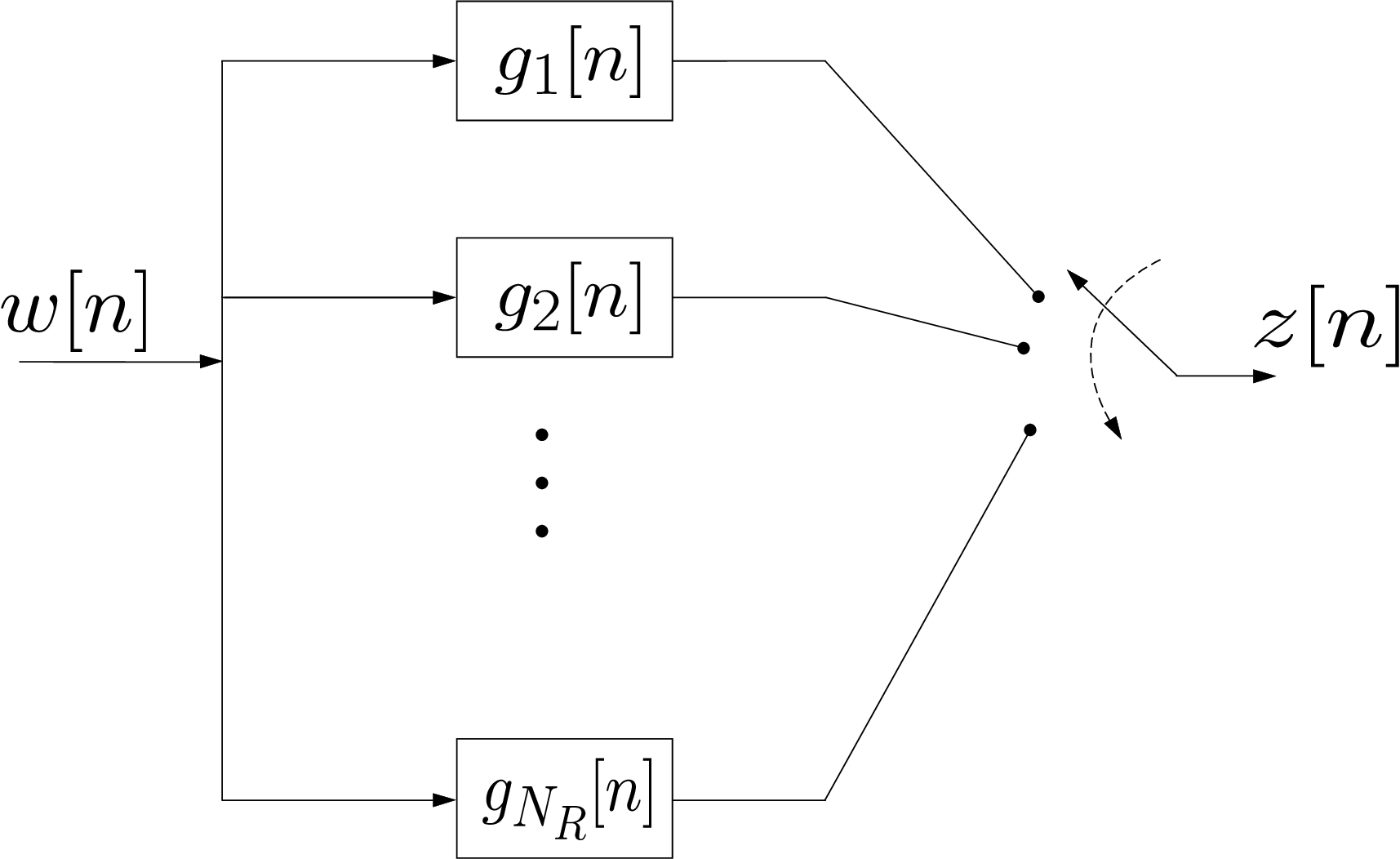}
           \caption{Nassar noise generation model: $z[n]$ is the result of sequential filtering of the zero mean unit variance white Gaussian input process $w[n]$ by a sequence of LTI filters $g_{i}[n]$.}
	\label{fig:Nassar_model}
        
\end{figure}

In this work, we consider OFDM transmission, so we divide each period of the noise into a specific number of OFDM symbols. In both Nassar's \cite{nassar2012cyclostationary} and FRESH filtering \cite{elgenedy2016cyclostationary} models, the noise statistics are periodic. Each period could be divided into $N_{R}$ time regions. However, in Nassar's model, each time region corresponds to the output of one LTI filter in a bank of parallel LTI filters, where the filters are selected sequentially as shown in Fig~\ref{fig:Nassar_model}. In FRESH filtering, noise is the summation of a bank of $K$ LTI filters. In both models, there is a correlation between time domain samples. In addition, as shown in Fig.~\ref{fig:noise-fresh-mimo}, the MIMO FRESH filtering model also includes the spatial correlation between noise samples in different links of the MIMO channel which are the phases of the electrical network. 

\begin{figure}
    \centering
    \includegraphics[width=0.4\textwidth]{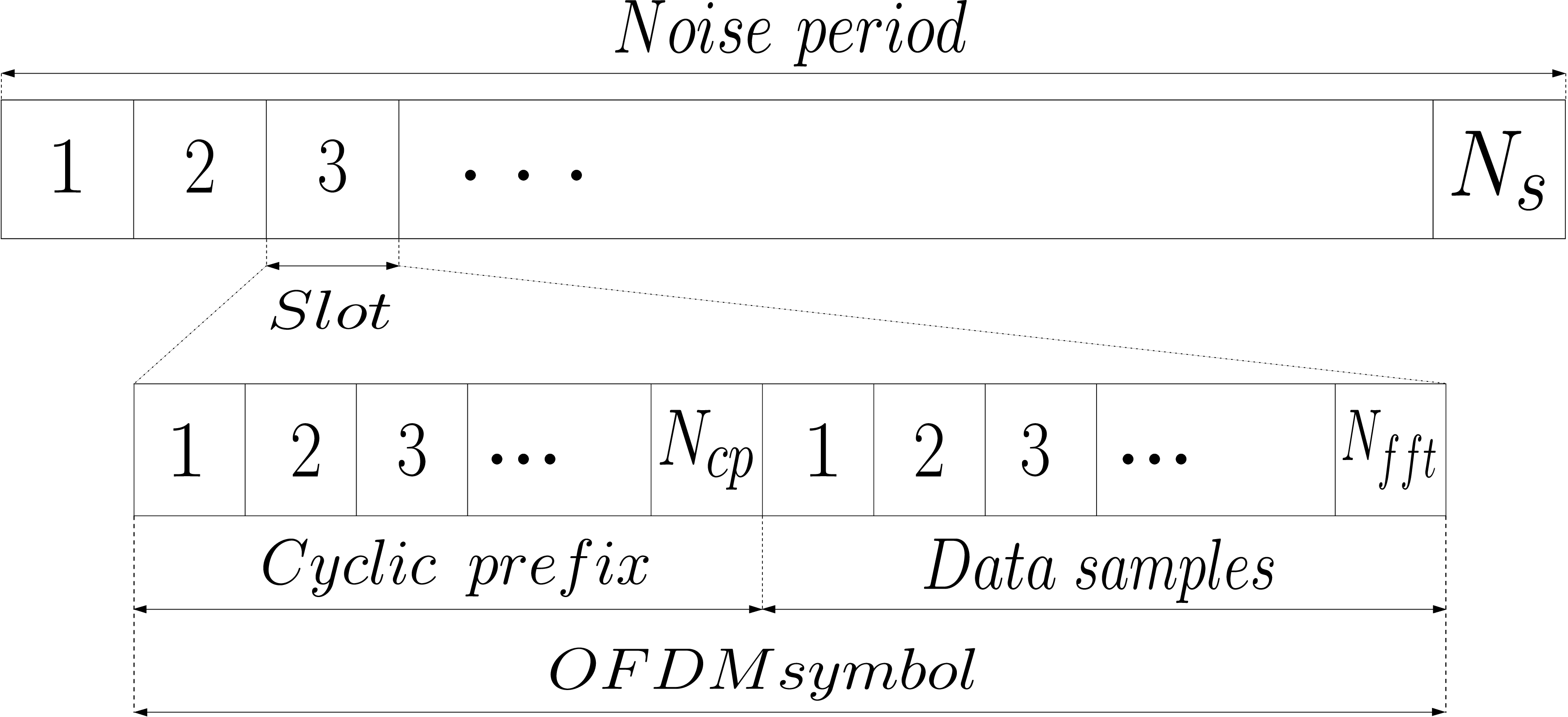}
    \caption{One period of the noise is divided into $N_{\text{s}}$ slots and each slot corresponds to one OFDM symbol with $N_\text{fft}+N_\text{cp}$ samples.}
    \label{fig:frame}
    \vspace{0.5 cm}
\end{figure}

In this section, similar to \cite{moaveninejad2019gaussian}, we assume that each period is divided into $N_{\text{s}}$ slots with each slot comprising a specific number of noise samples generated by the FRESH filter. Our goal is to show the results in accordance with the OFDM format proposed in IEEE 1901.2 standard \cite{standard}. To achieve this, we set the number of samples in each slot equal to the length of one OFDM symbol plus the cyclic prefix (CP), which is $N_\text{fft}+N_\text{cp}$ where $N_\text{cp}$ is equal to the length of the channel memory.
Figure~\ref{fig:frame} shows this partitioning. Note that, CP for data transmission refers to a copy of the last $N_\text{cp}$ samples inserted at the beginning of each OFDM symbol. Regarding CP in each slot of noise, $N_\text{cp}$ samples from the output of FRESH filter are considered to be superimposed to the CP of data signals. In other words, in this paper with regards to noise, $N_\text{cp}$ does not imply a copy of noise samples in each slot of the FRESH noise but it indicates the number of noise samples considered to be added to the CP of the transmitted data. Therefore, in each period with $N_{\text{smp}}$ samples there are $N_{\text{s}} =\frac{N_\text{smp}}{N_\text{fft}+N_\text{cp}}$ slots. After the Gaussianity test, each OFDM symbol may subsequently be divided into portions of $N_{\text{p}}+N_{\text{cp}}$ samples, such that the samples belonging to each portion appended by the CP are normally distributed. The length of each portion may be equal to the length of one OFDM symbol or to a fraction of it. 


\subsection{Gaussianity Test}
\label{sec:gaussianity_test}
As mentioned above, the noise samples in one period are divided into $N_{\text{s}}$ slots, each consisting of $N_\text{fft}+N_\text{cp}$ samples. In~\cite{moaveninejad2019gaussian}, it was shown that each slot belongs to one of the three different classes of noise, where each class is characterized by a different pdf. Although each individual sample is normally distributed in sequential periods, \cite{moaveninejad2019gaussian} shows that inside each period a sequence of consecutive samples belonging to the same class exhibits either an impulsive or a Gaussian pdf. In this paper, we seek to examine the extent to which the distribution of samples within each class approximates a Gaussian distribution. Keeping this issue in mind, we use a Gaussianity test to define temporal portions in each cycle inside which noise samples tend to become normally distributed. Towards this goal, we consider a further division of each OFDM symbol into portions, each containing $N_{\text{p}}=[\frac{N_{\text{fft}}+N_{\text{cp}}}{A}]-N_{\text{cp}}$ samples plus CP, with $A = 2^{\left[0, 1,\cdots, \log_2(\frac{N_{\text{fft}}+N_{\text{cp}}}{1+N_{\text{cp}}})\right]}$, for which the resulting distribution turns out to be normal. Consequently, each cycle is divided into $N_{\text{s}}\times A$ smaller portions each containing $N_{\text{p}}+N_\text{cp}$ samples. After the Gaussianity test is performed for all the possible divisions, the one with the largest $N_{\text{p}}$ is set as the length of each temporal portion in which noise samples can be considered as normally distributed. 

Each temporal portion $p$ with sequence of $N_{\text{p}}$ samples plus CP
\begin{equation}
z^{p}_\text{cp} = \left \{ z[n], z[n+1],\cdots, z[n+N_{\text{p}}+N_\text{cp}-1] \right \},
\end{equation}
is used to build a histogram $q^{p}_{z}(x)$ of the samples distribution. To test the Gaussianity of each temporal portion of the noise, we compute the Kullback-Leibler divergence (KLD) \cite{Thomas} between $q^{p}_{z}(x)$ and the Gaussian distribution
\begin{align}
g^{p}(x|\hat{\mu}_{z^{p}_\text{cp}}, \hat{\sigma}_{z^{p}_\text{cp}} ^{2}) = \frac{1}{\sqrt{2\pi \hat{\sigma}_{z^{p}_\text{cp}}^{2} }} \exp \left( -\frac{(x-\hat{\mu}_{z^{p}_\text{cp}})^{2}}{2\hat{\sigma}_{z^{p}_\text{cp}}^{2}} \right) ,
\end{align}
which is the pdf for random variable $X$ with $N_{\text{p}}+N_\text{cp}$ samples that are normally distributed with variance $\hat{\sigma}_{z^{p}_\text{cp}}^{2}$ and mean value $\hat{\mu}_{z^{p}_\text{cp}}$ estimated for $q^{p}_{z}(x)$. To estimate $\hat{\mu}_{z^{p}_\text{cp}}, \hat{\sigma}_{z^{p}_\text{cp}}^{2}$ for a certain portion of time, the samples corresponding to that portion are collected from $N_\text{period}$ cycles. Then, the mean and the variance are estimated for $N_\text{period}\times (N_{\text{p}}+N_\text{cp})$ samples corresponding to that specific portion. This is repeated for all $p=1, \cdots, N_{\text{s}}\times A$ portions in each cycle. The KLD between the pdfs $g^{p}(x|\hat{\mu}_{z^{p}_\text{cp}}, \hat{\sigma}_{z^{p}_\text{cp}} ^{2})=g^{p}(x)$ and $q^{p}_{z}(x)$ is
\begin{equation}
(D_{KL})^p \hspace{-.1cm}=\hspace{-.1cm} D_{KL}(q^{p}_{z}(x)||g^{p}(x)) \hspace{-.1cm}= \hspace{-.1cm}
\int_{-\infty}^{\infty} q^{p}_{z}(x) \log{\left(\frac{q^{p}_{z}(x)}{g^{p}(x)}\right)}dx.
\label{eq:KL}
\end{equation}
For each portion $p$, the $(D_{KL})^p$ is computed for $N_\text{itr}$ iterations and averaged over all $N_\text{itr}$ to set the value of $(D_{KL})^p$. The value of $D_{KL}$ measures how different $q^{p}_{z}(x)$ is from $g^{p}(x)$. Values of $ D_{KL}(q^{p}_{z}(x)||g^{p}(x))$ equal to $0$ indicate that $z^{p}_\text{cp}$ is Gaussian. Hence, we can define a Gaussianity threshold such that for different lengths of $N_{\text{p}}+N_\text{cp}$, if $(D_\text{KL})^p$ is below the threshold, then the corresponding $N_{\text{p}}$ is fixed as the length of temporal portions of noise with Gaussian behavior.

\subsection{Input-Output Model for MIMO System with LTI Channel}
\label{subsec:in-out-mimo}

In a SISO system, the $n$th sample of the received signal at the output of the LTI channel $h[n]$ is given by
\begin{align}
    y[n] = h[n]\ast x[n] + z[n],
    \label{formul:ys}
\end{align}
where $\ast$ denotes the discrete-time convolution. In case of a multipath channel with finite memory $L$, we have
\begin{align}
    y[n] = \sum_{l=0}^{L-1} h[l]x[n-l] + z[n],
    \label{eq:ys}
\end{align}
where $h[l]$$\,=\,$$\sum_{i=0}^{L-1} h_i \delta[l-i]$ and $\delta[i]$ is the Kronecker delta sequence. 

For an $M_{r}\times M_{t}$ MIMO system, let $h^{(r,t)}[n]$ denote the impulse response between the $t$th transmitter and the $r$th receiver. Therefore, \eqref{eq:ys} is extended as 
\begin{equation}
    y^{(r)}\hspace{-.05cm}[n] \hspace{-.1cm} = \hspace{-.15cm} \sum_{t=1}^{M_{t}}\hspace{-.1cm}\sum_{l=0}^{L\hspace{-.05cm}-\hspace{-.05cm}1}\hspace{-.1cm}h^{(r,t)}[l]x^{(t)}[n \hspace{-.05cm} - \hspace{-.05cm} l] \hspace{-.05cm} + \hspace{-.1cm} z^{(r)}[n],\ r \hspace{-.1cm} = \hspace{-.1cm} 1,\hspace{-.05cm}\cdots \hspace{-.06cm},M_r.
\label{eq:y_i}
\end{equation}
By using a more compact matrix notation, \eqref{eq:y_i} can be written as
\begin{align}
Y[n]=\sum_{l=0}^{L-1}H[l]X[n-l]+Z[n],
\label{eq:Yn}
\end{align}
where
\begin{equation}\hspace{-.1cm}
Y[n] \hspace{-.1cm}=\hspace{-.15cm}
\begin{bmatrix}\hspace{-.1cm}
y^{(1)}[n]\\y^{(2)}[n]\\ \vdots \\y^{(M_{r})}[n]\hspace{-.05cm}
\end{bmatrix}\hspace{-.1cm}
,X[n] \hspace{-.1cm}=\hspace{-.15cm}
\begin{bmatrix}\hspace{-.1cm}
x^{(1)}[n]\\x^{(2)}[n]\\ \vdots \\x^{(M_{t})}[n]\hspace{-.05cm}
\end{bmatrix}\hspace{-.1cm}
,Z[n] \hspace{-.1cm}=\hspace{-.15cm}
\begin{bmatrix}\hspace{-.1cm}
z^{(1)}[n]\\z^{(2)}[n]\\ \vdots \\z^{(M_{r})}[n]\hspace{-.05cm}
\end{bmatrix},
\label{eq:Yn,Xn,Zn}
\end{equation}
and, 
    \begin{align}
    H[l] =
\begin{bmatrix}
h^{(1,1)}[l] & h^{(1,2)}[l] &\cdots & h^{(1,M_{t})}[l]\\ 
h^{(2,1)}[l] & h^{(2,2)}[l] & \cdots & h^{(2,M_{t})}[l]\\ 
 \vdots& \vdots&\cdots &\vdots\\ 
 h^{(M_{r},1)}[l]&  h^{(M_{r},2)}[l] & \cdots &h^{(M_{r},M_{t})}[l] 
\end{bmatrix}.
\end{align}
By adding the CP to the transmitted signal, the linear convolution can be converted into a circular convolution. 
With the introduced matrix notation we can write
\begin{align}
    Y^{p}_\text{cp} & = H^{p}_\text{cp-cp} X^{p}_\text{cp}+Z^{p}_\text{cp}, 
    \label{eq:Yp_cp_cp_linear}
\end{align}
where 
{\footnotesize
\begin{equation}\hspace{-.15cm}
Y^{p}_\text{cp}\hspace{-.08cm}=\hspace{-.12cm}\begin{bmatrix}
Y[n\hspace{-.07cm}-\hspace{-.07cm}(L\hspace{-.07cm}-\hspace{-.07cm}1)] \\ \vdots \\ Y[n-1]\\ Y[n] \\ \vdots \\Y[n\hspace{-.05cm}+\hspace{-.05cm}N_{\text{p}}\hspace{-.09cm}-\hspace{-.05cm}1]
\end{bmatrix}\hspace{-.07cm},\hspace{-.02cm}
X^{p}_\text{cp}\hspace{-.08cm}=\hspace{-.12cm}\begin{bmatrix}
X[n\hspace{-.05cm}-\hspace{-.05cm}(L\hspace{-.07cm}-\hspace{-.07cm}1)] \\ \vdots\\ X[n-1] \\ X[n] \\ \vdots\\X[n\hspace{-.05cm}+\hspace{-.05cm}N_{\text{p}}\hspace{-.09cm}-\hspace{-.05cm}1]\\
\end{bmatrix}\hspace{-.07cm},\hspace{-.02cm}
Z^{p}_\text{cp}\hspace{-.08cm}=\hspace{-.12cm}\begin{bmatrix}
Z[n\hspace{-.05cm}-\hspace{-.05cm}(L\hspace{-.07cm}-\hspace{-.07cm}1)]\\ \vdots \\Z[n-1] \\ Z[n] \\ \vdots\\Z[n\hspace{-.05cm}+\hspace{-.05cm}{N_{\text{p}}}\hspace{-.09cm}-\hspace{-.05cm}1] 
\end{bmatrix},
\end{equation}}
{\footnotesize
\begin{equation}\hspace{-.15cm}
H^{p}_\text{cp-cp}=
\begin{bmatrix}
\hspace{-.05cm}H[0] & \hspace{-.2cm}0 & \hspace{-.2cm}\cdots & \hspace{-.2cm}\cdots & \hspace{-.2cm}\cdots & \hspace{-.2cm}\cdots & \hspace{-.2cm}\cdots & \hspace{-.2cm}0\\ 
\hspace{-.05cm}H[1] & \hspace{-.2cm}H[0] & \hspace{-.2cm}0 & \hspace{-.2cm}\cdots  & \hspace{-.2cm}\cdots & \hspace{-.2cm}\cdots & \hspace{-.2cm}\cdots & \hspace{-.2cm}0\\
\hspace{-.1cm}\vdots & \hspace{-.2cm}\ddots & \hspace{-.2cm}\ddots & \hspace{-.2cm}\ddots & \hspace{-.2cm}& \hspace{-.2cm}& \hspace{-.2cm}&\hspace{-.2cm}\vdots\\ 
\hspace{-.05cm}H[L\hspace{-.1cm}-\hspace{-.1cm}2] & \hspace{-.2cm}\cdots & \hspace{-.2cm}\cdots & \hspace{-.2cm}H[0] & \hspace{-.2cm}0 & \hspace{-.2cm}\cdots & \hspace{-.2cm}\cdots & \hspace{-.2cm}\vdots \\ 
\hspace{-.05cm}H[L\hspace{-.1cm}-\hspace{-.1cm}1] & \hspace{-.2cm}H[L\hspace{-.1cm}-\hspace{-.1cm}2] & \hspace{-.2cm}\cdots & \hspace{-.2cm}\cdots &  \hspace{-.2cm}H[0] & \hspace{-.2cm}0 &\hspace{-.2cm}\cdots & \hspace{-.1cm}\vdots \\ 
\hspace{-.05cm}0 & \hspace{-.2cm}H[L\hspace{-.1cm}-\hspace{-.1cm}1] & \hspace{-.2cm}\cdots & \hspace{-.2cm}\cdots & \hspace{-.2cm}\cdots & \hspace{-.2cm}H[0]  & \hspace{-.2cm}0 & \hspace{-.2cm}\vdots\\ 
\hspace{-.05cm}\vdots & \hspace{-.2cm}\ddots & \hspace{-.2cm}\ddots & \hspace{-.2cm}\ddots & \hspace{-.2cm} &  & \hspace{-.2cm}\ddots &\hspace{-.2cm}\vdots\\ 
\hspace{-.05cm}0 & \hspace{-.2cm}\cdots & \hspace{-.2cm}0 & \hspace{-.2cm}H[L\hspace{-.1cm}-\hspace{-.1cm}1] & \hspace{-.2cm}\cdots & \hspace{-.2cm}\cdots & \hspace{-.2cm}\cdots & \hspace{-.2cm}H[0]
\end{bmatrix}.
\label{eq:Hs_cp-cp}
\end{equation}}
By adding the CP, the data vector transmitted from each transmitter during one portion will contain $N_{\text{p}}+N_\text{cp}$ samples. Therefore, signals transmitted from $M_{t}$ antennas are represented by $X^{p}_\text{cp}$ which is an $({M_{t}}\cdot(N_{\text{p}}+N_\text{cp})) \times 1$ vector. Similarly, $Y^{p}_\text{cp}$ and $Z^{p}_\text{cp}$ correspond to the received signals and noise vectors with the same number of samples per Gaussian portion at $M_{r}$ receivers. The matrix $H^{p}_\text{cp-cp}$ represents the convolutive intersymbol interference (ISI) linear channel matrix with an additional $N_\text{cp}$ columns and $N_\text{cp}$ rows added to its $N_{\text{p}}$ columns and rows, respectively. The result is an $(M_{r}\cdot(N_{\text{p}}+N_\text{cp})) \times (M_{t}\cdot(N_{\text{p}}+N_\text{cp})) $ MIMO channel matrix experienced by $X^{p}_\text{cp}$ \cite{wilzeck2008antenna} over one portion of time. Accordingly, each column of the matrix (\ref{eq:Hs_cp-cp}) for channel impulse response contains only $L$ non-zero components and is padded with an appropriate number of zeros equal to $M_{t} \cdot (N_{\text{p}}+N_\text{cp}-L)$.

The ISI extends over the first $L-1$ received samples which is ignored by the corresponding receiver. Therefore, for simplicity, here the output is assumed as the time reference, such that in presence of cyclic prefix, input time $n_{v}$ and output time $n_{u}$ are $(n_{v}, n_{u}) \in \{ -(L-1), \cdots , -1, 0, 1, \cdots , N_{\text{p}} - 1 \}$. Subsequently, the ISI is removed from each received signal vector by considering only the output over the time interval $n_{u} \in \left \{ 0, 1,\cdots,N_{\text{p}}-1 \right \}$. 
Therefore, (\ref{eq:Yp_cp_cp_linear}) can be expressed as follows
\begin{align}
    Y^{p}= H^{p}_\text{cp}X^{p}_\text{cp}+Z^{p},
    \label{eq:Yp_cp_linear}
\end{align} 
\begin{align}
Y^{p} =\begin{bmatrix}
Y[n] \\ \vdots \\Y[n+N_{\text{p}}-1] 
\end{bmatrix},
Z^{p} = \begin{bmatrix}
Z[n] \\ \vdots\\Z[n+{N_{\text{p}}}-1]
\end{bmatrix}.
\label{eq:Yss,Zss}
\end{align}
By removing the first $L-1$ rows, $Y^{p}_\text{cp}$, $Z^{p}_\text{cp}$ are replaced with $Y^{p}$, $Z^{p}$. Keeping this assumption in mind, channel matrix $H^{p}_\text{cp}$ is defined as
{\footnotesize
\begin{equation}\hspace{-.15cm}
H^{p}_\text{cp}=
\begin{bmatrix}
\hspace{-.05cm}H[L\hspace{-.1cm}-\hspace{-.1cm}1] & \hspace{-.2cm}H[L\hspace{-.1cm}-\hspace{-.1cm}2] & \hspace{-.2cm}\cdots & \hspace{-.2cm}H[1] &  \hspace{-.2cm}H[0] & \hspace{-.2cm}0 &\hspace{-.2cm}\cdots & \hspace{-.2cm}0 \\
\hspace{-.05cm}0 & \hspace{-.2cm}H[L\hspace{-.1cm}-\hspace{-.1cm}1] & \hspace{-.2cm}\cdots & \hspace{-.2cm}\cdots & \hspace{-.2cm}H[1] & \hspace{-.2cm}H[0]  & \hspace{-.2cm}0 & \hspace{-.2cm}\vdots\\
\vdots &  & \ddots &  & &  & \ddots &\vdots\\
\hspace{-.05cm}0 & \hspace{-.2cm}\cdots & \hspace{-.2cm}0 & \hspace{-.2cm}H[L\hspace{-.1cm}-\hspace{-.1cm}1] & \hspace{-.2cm}\cdots & \hspace{-.2cm}\cdots & \hspace{-.2cm}H[1] & \hspace{-.2cm}H[0]
\end{bmatrix}.
\label{eq:Hs_cp}
\end{equation}}%
The number of columns in $H^{p}_\text{cp}$ are the same as in $H^{p}_\text{cp-cp}$. Hence, $H^{p}_\text{cp}$ is an $(M_{r} \cdot N_{\text{p}}) \times (M_{t} \cdot (N_{\text{p}}+N_\text{cp})) $ MIMO channel linear convolution matrix. The elements of $H^{p}_\text{cp}$ are defined as $(H^{p}_\text{cp})_{(n_{u},n_{v})}= H[l]_{l=n_{u}-n_{v}}$ if $0 \leq n_{u}-n_{v} < L$ and $(H^{p}_\text{cp})_{(n_{u},n_{v})}=0$, otherwise. 

Models \eqref{eq:Yp_cp_cp_linear} and \eqref{eq:Yp_cp_linear} represent linear convolution, while for the circular convolution, \eqref{eq:Yp_cp_linear} is equivalent to
\begin{align}
    Y^{p} = {H^{p}}X^{p}+Z^{p},
    \label{eq:Yp_circular}
\end{align} 
where ${H^{p}}$ is an $(M_{r} \cdot N_{\text{p}}) \times (M_{t} \cdot N_{\text{p}}) $ MIMO circular channel matrix given by
{\footnotesize
\begin{equation}\hspace{-.15cm}
{H^{p}}=
\begin{bmatrix}
\hspace{-.05cm}H[0] & \hspace{-.2cm}0 & \hspace{-.2cm}\cdots & \hspace{-.2cm}H[L\hspace{-.1cm}-\hspace{-.1cm}1] &\hspace{-.2cm}\cdots & \hspace{-.2cm}H[1] \\
\hspace{-.05cm}H[1] & \hspace{-.2cm}H[0] & \hspace{-.2cm}0 & \hspace{-.2cm}\cdots  & \hspace{-.2cm}H[L\hspace{-.1cm}-\hspace{-.1cm}1]  & \hspace{-.2cm}\vdots \\
\hspace{-.05cm}\vdots & \hspace{-.2cm}\ddots & \hspace{-.2cm}H[0] & \hspace{-.2cm}0 & \hspace{-.2cm}\cdots &  H[L\hspace{-.1cm}-\hspace{-.1cm}1] \\
\hspace{-.05cm}H[L\hspace{-.1cm}-\hspace{-.1cm}1] & \hspace{-.2cm}\cdots & \hspace{-.2cm}H[1] & \hspace{-.2cm}H[0] & \hspace{-.2cm}0 & \hspace{-.2cm}0\\
\hspace{-.05cm}0 & \hspace{-.2cm}\ddots &  & \hspace{-.2cm}\ddots & \hspace{-.2cm}H[0] &  \hspace{-.2cm}\vdots\\
\hspace{-.05cm}\vdots & \hspace{-.2cm}\cdots & \hspace{-.2cm}H[L\hspace{-.1cm}-\hspace{-.1cm}1]  & \hspace{-.2cm}\cdots & \hspace{-.2cm}H[1] & \hspace{-.2cm}H[0]
\end{bmatrix}.
\end{equation}}
It is worth noting that, in addition to compensating the effects of ISI, CP also serves to implement single-tap equalization and the interested reader is referred to~\cite{peng2010communications} for further insight.

\subsection{Noise Spatio-Temporal Whitening}
\label{subsec:noise temporal whitening}
As part of the capacity derivation in presence of correlated and colored noise, it is necessary to perform noise whitening for each Gaussian portion to eliminate the correlation between noise samples. Consequently, in the MIMO system, the problem of spatially and temporally correlated noise vectors $Z^{p}$, is transformed to multivariate zero-mean white Gaussian noise (WGN) $W^{p}$. This means that, after whitening, each portion of the noise becomes Gaussian with correlation matrix $R_{W^{p} W^{p}}$ equal to identity matrix. Note that, each period consists of $N_{\text{s}}\times A$ portions of the noise, hence after whitening there are $N_{\text{s}}\times A$ Gaussian portions with different variances.

For a MIMO system with $M_{r}$ receivers, the noise vector $Z^{p}$ for each temporal portion with $N_{\text{p}}$ samples is an $(M_{r} \cdot N_{\text{p}}) \times 1$ vector obtained by substituting (\ref{eq:Yn,Xn,Zn}) as $Z[n]$ inside (\ref{eq:Yss,Zss}) which gives 
\begin{equation}\hspace{-.3cm}
Z^{p}\hspace{-.15cm}=\hspace{-.1cm}\begin{bmatrix}\hspace{-.03cm}
\begin{bmatrix}
z^{1}\hspace{-.05cm}[n] 
\hspace{-.05cm} \cdots \hspace{-.05cm}
z^{M_{r}}\hspace{-.05cm}[n]
\end{bmatrix}
\hspace{-.07cm} \cdots \hspace{-.07cm}
\begin{bmatrix}
z^{1}\hspace{-.05cm}[n\hspace{-.1cm}+\hspace{-.1cm}N_{\text{p}}\hspace{-.17cm}-\hspace{-.1cm}1]
\hspace{-.05cm} \cdots \hspace{-.05cm}
z^{M_{r}}\hspace{-.05cm}[n\hspace{-.1cm}+\hspace{-.1cm}N_{\text{p}}\hspace{-.17cm}-\hspace{-.1cm}1]
\end{bmatrix}\hspace{-.03cm}
\end{bmatrix}\hspace{-.1cm}^{T}.
\end{equation} 
The autocorrelation matrix $R_{Z^{p} Z^{p}}$ for each temporal portion of the noise in this MIMO system is an $(M_{r} \cdot N_{\text{p}}) \times (M_{r} \cdot N_{\text{p}})$ matrix which is obtained from
\begin{equation}
\label{eq:Rzs}
R_{Z^{p} Z^{p}} = E \{ Z^{p} [n:n+N_{\text{p}}-1] (Z^{p}[n:n+N_{\text{p}}-1])^{\dagger} \},
\end{equation}
where $E\left \{ \cdot \right \}$ is the expectation operator and the superscript $^\dagger$ denotes the complex conjugate transpose.

The temporal and spatial decorrelation is realized by multiplying the noise vector $Z^{p}$ by the whitening matrix $R_{Z^{p} Z^{p}}^{-\frac{1}{2}}$
\begin{equation}
    W^{p} = R_{Z^{p} Z^{p}}^{-\frac{1}{2}} Z^{p}.
\end{equation}
The square root matrix $R_{Z^{p} Z^{p}}^{-\frac{1}{2}}$ is computed using the Cholesky factorization \cite{Cholesky_decomp} as
\begin{align}
    R_{Z^{p} Z^{p}} &= R_{Z^{p} Z^{p}}^{\frac{1}{2}} \left(R_{Z^{p} Z^{p}}^{\frac{1}{2}}\right)^\dagger = L L^{\dagger}, \label{eq:R_Zp}\\
    R_{Z^{p} Z^{p}}^{-1} &= \left(R_{Z^{p} Z^{p}}^{-\frac{1}{2}}\right)^\dagger R_{Z^{p} Z^{p}}^{-\frac{1}{2}} = ({L^{-1}})^{\dagger} L^{-1}\label{eq:R'_Zp},
\end{align}
where $L$ is the lower triangular matrix from Cholesky factorization.

The decorrelation can be verified from the autocorrelation matrix of the whitened noise
\begin{align}
    R_{W^{p} W^{p}} &= E \{W^{p}  ({W^{p}})^{\dagger} \}\\
            &= E \left\{ R_{Z^{p} Z^{p}}^{-\frac{1}{2}} Z^{p} ({Z^{p}})^{\dagger} \left(R_{Z^{p} Z^{p}}^{-\frac{1}{2}}\right)^\dagger \right\} \nonumber \\
            &= R_{Z^{p} Z^{p}}^{-\frac{1}{2}} R_{Z^{p} Z^{p}} \left(R_{Z^{p} Z^{p}}^{-\frac{1}{2}}\right)^\dagger = I, \nonumber 
\end{align}
where $I$ is the identity matrix.

The noise pre-whitening needs to be performed for the received signal $Y^{p}$. Henceforth, $W^{p}$ is referred to as an AWGN that is independent across the receivers. Moreover, the effect of noise whitening is lumped with the channel so ${\widehat{H}^{p}_\text{cp}}$ and ${\widehat{H}^{p}}$ represent composite linear channel matrix and composite circular channel matrix, respectively. Therefore, $H^{p}$ loses the circulant structure after whitening and is changed to ${\widehat{H}^{p}}$. The received signal after whitening is obtained from
\begin{align}
   {\widehat{Y}^{p}} & = R_{Z^{p} Z^{p}}^{-\frac{1}{2}}Y^{p}. 
\end{align}
Using $Y^{p}$ from \eqref{eq:Yp_cp_linear} we have
\begin{align}
    { \widehat{Y}^{p}} & = R_{Z^{p} Z^{p}}^{-\frac{1}{2}} (H^{p}_\text{cp} X^{p}_\text{cp} + Z^{p}) \nonumber \\
    & = {\widehat{H}_\text{cp}^{p}} X^{p}_\text{cp} + W^{p}.
        \label{eq:Y's_linear}
\end{align}
Equivalently, using \eqref{eq:Yp_circular} for $Y^{p}$ gives
\begin{align}
  { \widehat{Y}^{p}} & = R_{Z^{p} Z^{p}}^{-\frac{1}{2}} (H^{p} X^{p} + Z^{p}) \nonumber \\
   & = {\widehat{H}^{p}} X^{p} + W^{p}.
   \label{eq:Y_white}
\end{align}

\subsection{Capacity of MIMO for Each Temporal Portion with Gaussian Behavior}
\label{subsec:capacity}

Shannon's channel capacity is given by the maximum of the mutual information between transmitter and receiver vectors over all distributions of the input that satisfy the average power constraint
\begin{align}\label{eq:max_mutual_xy}
    C^{p} & = \max_{\text{tr}(R_{X^{p} X^{p}}) \leq \text{P}} {I({\widehat Y}^{p};X^{p})},
\end{align}
where P is the average power available at the transmitter, $\text{tr}(\cdot)$ denotes the trace of the matrix, and
\begin{align}
    I({\widehat Y}^{p};X^{p}) & = h({\widehat Y}^{p})-h({\widehat Y}^{p}|X^{p}) \nonumber \\
    & =  h({\widehat Y}^{p})- h( W^{p}),
    \label{eq:mutual_xy}
\end{align}
where $h(Q)$ denotes the differential entropy of random variable $Q$ and the second equality comes from the perfect channel knowledge at the receiver. The differential entropies for the zero mean random variables ${\widehat Y}^{p}$ and $W^{p}$ with known correlation matrices $R_{{W^{p}}{W^{p}}}$ and
\begin{align}
     R_{{\widehat Y^{p}} {\widehat Y^{p}}} & = E \{ {\widehat Y}^{p} ({\widehat Y^{p}})^\dagger \} \\
     & = {\widehat H^{p}} R_{ X^{p} X^{p} } ({\widehat H^{p}})^{\dagger} + R_{ {W^{p}} {W^{p}} },
\end{align}
are upper-bounded as
\begin{align}
    h({\widehat Y}^{p}) & \leq \frac{1}{2} \log_{2}\left [ (2 \pi e)^{n} \left | R_{{\widehat Y}^{p} {\widehat Y}^{p}} \right |\right ],\label{eq:entropy_y}\\
     h( W^{p}) & \leq  \frac{1}{2}\log_{2}\left [ (2 \pi e)^{n} \left | R_{{W^{p}}{W^{p}}} \right |\right ],\label{eq:entropy_w}
\end{align}
where $|\cdot|$ denotes the determinant operation.

The inequalities in \eqref{eq:entropy_y} and \eqref{eq:entropy_w} hold with equality when ${\widehat Y}^{p}$ and $W^{p}$ are Gaussian random vectors \cite{Thomas}. Accordingly, since $W^{p}$ is normally distributed, for the given composite channel matrix ${\widehat H^{p}}$, $h({\widehat Y}^{p})$ is maximized when $X^{p}$ is also a zero mean Gaussian random variable. Under this assumption, the supremum of \eqref{eq:max_mutual_xy} is achieved and it becomes
\begin{align}
   C^{p} & \hspace{-.05cm}=\hspace{-.05cm} \log_{2} \hspace{-.05cm} \left |  I_{M_{r} \cdot N_{\text{p}}} \hspace{-.08cm} + \hspace{-.08cm} R_{W^{p} W^{p}}^{-1} {\widehat H^{p}} R_{X^{p} X^{p}} ({\widehat H^{p}})^\dagger \right |.
\end{align}

\subsubsection{Unknown Channel State at the Transmitter}
When the channel/noise state is unknown at the transmitter, the transmit power in each temporal portion $P_{X^{p}}$ is shared equally between $\{ x[n] \}_{n=0:N_{\text{p}}+N_\text{cp}-1}$ signals transmitted from each of the $M_{t}$ transmitters. This gives rise to the correlation matrix $R_{X^{p} X^{p}} $ fixed to $ \varepsilon_{p} I_{M_{t} \cdot (N_{\text{p}}+N_{\text{cp}})}$
, where $\varepsilon_{p}$ is the power at each time instance $n=0, \cdots, N_{\text{p}}+N_\text{cp}-1$ of a specific portion $p$ given by
\begin{align}
    \varepsilon_{p} = \frac{P_{X^{p}}}{M_{t}(N_{\text{p}}+N_\text{cp})}.
\end{align}
Considering this power distribution and whitened Gaussian noise with $R_{W^{p} W^{p}} = I_{M_{r} \cdot N_{\text{p}}} $, the capacity can be expressed as
\begin{align}
    C^{p} & = \log_{2}\left | I_{M_{r} \cdot N_{\text{p}}} +  \varepsilon_{p} {\widehat H^{p}} ({\widehat H^{p}})^{\dagger}  \right |,
    \label{eq:cap_time_slot}
\end{align}
in bits per temporal portion of the MIMO channel.
Following the eigen-decomposition ${\widehat H^{p}} ({\widehat H^{p}})^\dagger = {\widehat Q^{p}} \Lambda^{p} ({\widehat Q^{p}})^\dagger$,  \eqref{eq:cap_time_slot} can be expressed as
\begin{align}
    C^{p} & = \log_{2}\left | I_{M_{r} \cdot N_{\text{p}}} +  \varepsilon_{p} {\widehat Q^{p}} \Lambda^{p} ({\widehat Q^{p}})^{\dagger}  \right |\nonumber\\
    & = \log_{2}\left | {\widehat Q^{p}} \left( I_{M_{r} \cdot N_{\text{p}}} +  \varepsilon_{p} \Lambda^{p} \right ) ({\widehat Q^{p}})^{\dagger}  \right |,
    \label{eq:cap_time_q}
\end{align}
and since $[\det(({\widehat Q^{p}})^\dagger)]^{*} = \det({\widehat Q^{p}}) = 1$, then
\begin{align}
    C^{p} = \sum\limits_{i=1}^{m} \log_{2} \left ( 1 +  \varepsilon_{p} \delta^{i} \right ),
    \label{eq:cap_time_gamma2}
\end{align}
where $m$ is the rank of the matrix (${\widehat H^{p}} ({\widehat H^{p}})^{\dagger}$) and $\{\delta^{i}\}_{i=1 : m}$ are its eigenvalues.
\subsubsection{Known Channel State at the Transmitter}
When both the channel and noise statistics are constant and known at the transmit side, the capacity can be improved by applying the waterfilling algorithm \cite{Thomas}, which allocates more power to the channel with higher signal-to-noise ratio (SNR). Furthermore, if an SNR value is below a given threshold $\mu$, the corresponding channel must not be used, that is, no power allocated to it. In this case, constant value for the energy of transmitted signal $\varepsilon_{p}$ is replaced by $\varepsilon^{i}_{p}$, which is the optimized power allocated to that signal based on waterfilling
\begin{align}
   C^{p} = \sum\limits_{i=1}^{m} \log_{2} \left ( 1 + \varepsilon^{i}_{p} \delta^{i} \right ).
  \label{eq:cap_chanel_known}
\end{align}

\subsection{Capacity of MIMO-OFDM}
\label{subsec:cap_mimo_ofdm}
The OFDM scheme converts communication over a multipath channel into communication over parallel narrowband subcarriers in the frequency domain. The transformation between the time and the frequency domain is done by means of the discrete Fourier transform (DFT) and inverse DFT (IDFT) operations that are implemented efficiently by using the fast Fourier transform (FFT) and inverse FFT (IFFT), respectively.
\subsubsection{Before Noise Whitening}
By applying the Fourier transform on the circular convolution in (\ref{eq:Yp_circular}), the representation of the input-output relationship is
\begin{align}
   \mathcal{Y}^{p} & = \text{FFT} \{ H^{p} X^{p} + Z^{p} \} \nonumber\\
    & = \mathcal{H}^{p} \mathcal{X}^{p} + \mathcal{Z}^{p},
\end{align}
where $\mathcal{H}^{p}$ is the Fourier transform of the circular matrix $H^{p}$. The received symbol on each subcarrier $k$ in the temporal portion $p$ is given by:
\begin{align}
{\mathcal Y}^{p} &= 
\begin{bmatrix}
 {\mathcal Y}_{k}, \cdots, {\mathcal Y}_{k+N_{\text{p}}-1}
\end{bmatrix}^{T}\\
   {\mathcal Y}_{k} &= {\mathcal H}_{k} \mathcal X_{k} + \mathcal Z_{k},
\end{align}
where
\begin{equation}\hspace{-.1cm}
{\mathcal Y}_{k} \hspace{-.1cm}=\hspace{-.15cm}
\begin{bmatrix}\hspace{-.1cm}
y^{(1)}_{k}\\y^{(2)}_{k}\\ \vdots \\y^{(N_{r})}_{k}\hspace{-.05cm}
\end{bmatrix}\hspace{-.1cm}
,{\mathcal X}_{k} \hspace{-.1cm}=\hspace{-.15cm}
\begin{bmatrix}\hspace{-.1cm}
x^{(1)}_{k}\\x^{(2)}_{k}\\ \vdots \\x^{(N_{t})}_{k}\hspace{-.05cm}
\end{bmatrix}\hspace{-.1cm}
,{\mathcal Z}_{k} \hspace{-.1cm}=\hspace{-.15cm}
\begin{bmatrix}\hspace{-.1cm}
z^{(1)}_{k}\\z^{(2)}_{k}\\ \vdots \\z^{(N_{r})}_{k}\hspace{-.05cm}
\end{bmatrix},
\end{equation}
\begin{align}
{\mathcal H}_{k} &= 
\begin{bmatrix}
h^{(1,1)}_{k} & h^{(1,2)}_{k} &\cdots & h^{(1,M_{t})}_{k}\\ 
h^{(2,1)}_{k} & h^{(2,2)}_{k} & \cdots & h^{(2,M_{t})}_{k}\\ 
 \vdots& \vdots&\cdots &\vdots\\ 
 h^{(M_{r},1)}_{k}&  h^{(M_{r},2)}_{k} & \cdots &h^{(M_{r},M_{t})}_{k} 
\end{bmatrix}.
\end{align}
In these matrices, $y_{k}^{(r)}$ and $z_{k}^{(r)}$ are the received symbol and noise sample on the $k$th subcarrier and $r$th receiver, respectively. Similarly, $x_{k}^{(t)}$ is the transmitted symbol on the $k$th subcarrier and $t$th transmitter. The frequency response of the channel between transmitter $t$ and receiver $r$ is defined as
\begin{align}
    {h}_{k}^{(r,t)} &= \sum_{l=0}^{L-1}h^{(r,t)}[l] e^{{-j2\pi kl}/{N_{\text{p}}}}.
\end{align}
Channel capacity in one portion of time when noise has Gaussian behavior is already defined in (\ref{eq:cap_time_slot}) after the whitening of the noise in the time domain. This capacity could be obtained also in the frequency domain using OFDM and without whitening of the noise as follows
\begin{align}
    C^{p} & = \sum\limits_{k=1}^{m} \log_{2}\left | I_{M_{r}} + \varepsilon_{p}{\mathcal H}_{k} {\mathcal H}_{k}^\dagger  \right |.
    \label{eq:cap_freq}
\end{align}
Denoting $\mathcal F^{p}$ as the unitary FFT matrix over $N_{\text{p}}$ samples in one portion of time, then $X^{p}$ is the transmitted signal after passing through IFFT at transmitter side  $X^{p}= {(\mathcal F^{p})}^{\dagger} \mathcal X^{p}$. Furthermore, the circular matrix can be written as $H^{p}=  {(\mathcal F^{p})}^{\dagger} \Lambda^{p}   {\mathcal F^p}$. As a result, the FFT of the received signal ${Y^{p}}$ is given as:
\begin{align}
    \text{FFT} \{ {Y}^{p} \} &=  {\mathcal F^p} {Y^{p}}\\
     &=  {\mathcal F^p} ( H^{p} X^{p} + Z^{p}) \\
     &=  {\mathcal F^p} ( {(\mathcal F^p)}^{\dagger} \Lambda^{p}   {\mathcal F^p}  {(\mathcal F^p)}^{\dagger} \mathcal X^{p} + Z^{p}) \\
     &= \Lambda^{p} \mathcal X^{p} +  {\mathcal F^p} Z^{p}.
     \label{eq:fft_circular}
\end{align}
By adding a cyclic prefix and using the IFFT/FFT at transmitter and receiver sides, OFDM decomposes an ISI channel $H^{p}$ into $N_{\text{p}}$ orthogonal subchannels $\Lambda^{p}$, the knowledge of the channel matrix ${\mathcal H}^{p}$ is not needed for this decomposition. Therefore, the channel capacity in one temporal portion in (\ref{eq:cap_time_gamma2}) is equivalent to (\ref{eq:cap_freq_gamma2}) in the frequency domain
\begin{align}
    C^{p} = \sum\limits_{i=1}^{m} \log_{2} \left ( 1 +  \varepsilon_{p} \Lambda^{i} \right ).
    \label{eq:cap_freq_gamma2}
\end{align}
Similar to the time domain, if the channel and noise are known at the transmit side, $\varepsilon_{p}$ could be replaced by $\varepsilon^{i}_{p}$.
\subsubsection{After Noise Whitening}
We analyzed the capacity of each portion before applying whitening and decorrelating the noise samples. However, as shown in (\ref{eq:Y_white}), after the whitening of the noise, the circular channel matrix $H^p$ in (\ref{eq:Yp_circular}) is changed to a composite channel ${\widehat{H}^{p}}$ in (\ref{eq:Y_white}) and loses its circular properties. This means that, after applying FFT to the channel matrix it does not decompose to orthogonal subchannels $\Lambda^{p}$, as in OFDM systems. Therefore, \eqref{eq:fft_circular} for circular channel $H^p$ does not hold for the composite channel ${\widehat{H}^{p}}$. Hence, using circular channel does not help in calculation of capacity. Therefore, we may apply the FFT either to (\ref{eq:Y's_linear}), with composite linear channel matrix, or (\ref{eq:Y_white}), with composite circular channel matrix.
After applying the FFT to (\ref{eq:Y_white}), the channel capacity becomes
\begin{align}
    C^{p} & = \log_{2}\left | I_{M_{r} \cdot N_{\text{p}}} +  \varepsilon_{p} {\widehat {\mathcal H}^{p}} ({\widehat {\mathcal H}^{p}})^{\dagger} \right |.
    \label{eq:cap_time_slot_fft}
\end{align}
Here, $ {\widehat {\mathcal H}^{p}}$ is the FFT of the composite channel matrix $ {\widehat {H}^{p}}$. Given the eigenvalue decomposition, capacity is obtained in a similar way to that in \eqref{eq:cap_time_gamma2}. Consequently, for the case when the channel state is known at the transmitter, capacity after FFT is obtained in the same way as in (\ref{eq:cap_chanel_known}). 

\section{Numerical Evaluation}
\label{sec:numerical_evaluation}
In this section, we numerically evaluate a lower bound to the capacity we derived for the MIMO NB-PLC system with spatio-temporal correlated noise for the measured noise samples and an approximation of the capacity for the noise generated by the FRESH filtering model. The capacity is computed by dividing each period of time into different portions in which cyclostationary correlated noise exhibits Gaussian behavior. In this work, we set the length of each portion of the noise equal to the length of one OFDM symbol. Accordingly, we fix an overall threshold for Gaussianity that must be met by all slots. Finally, using the auxiliary-channel lower bound theorem from \cite{simulation-based}, we approximate our channel with a Gaussian channel and compute a lower bound to the capacity of the system. It is clear that by shortening the noise portions in time and moving towards a normal distribution, the bound approaches the actual capacity.

\subsection{System Model}
Based on the IEEE 1901.2 standard \cite{standard}, each OFDM symbol consists of $N_\text{fft}$$\,=\,$$256$ data samples. In this system, $L = 65$ taps is the length of the LTI channel with finite impulse response (FIR) and consequently $N_\text{cp} = 64$. Therefore, $N_\text{fft}+N_\text{cp}= 320$ samples pass through each input of the $2\times2$ MIMO channel. Under this condition, the linear convolution of the input sequence transmitted over the channel shown in (\ref{eq:Yp_cp_cp_linear}) and (\ref{eq:Yp_cp_linear}) is converted into a circular convolution as shown in (\ref{eq:Yp_circular}). The cyclostationary noise is added to the signal in the $2\times2$ MIMO NB-PLC channel. Finally, the CP is removed from the received signal. 



\begin{figure}
	\centering
	\includegraphics[width=1\columnwidth]{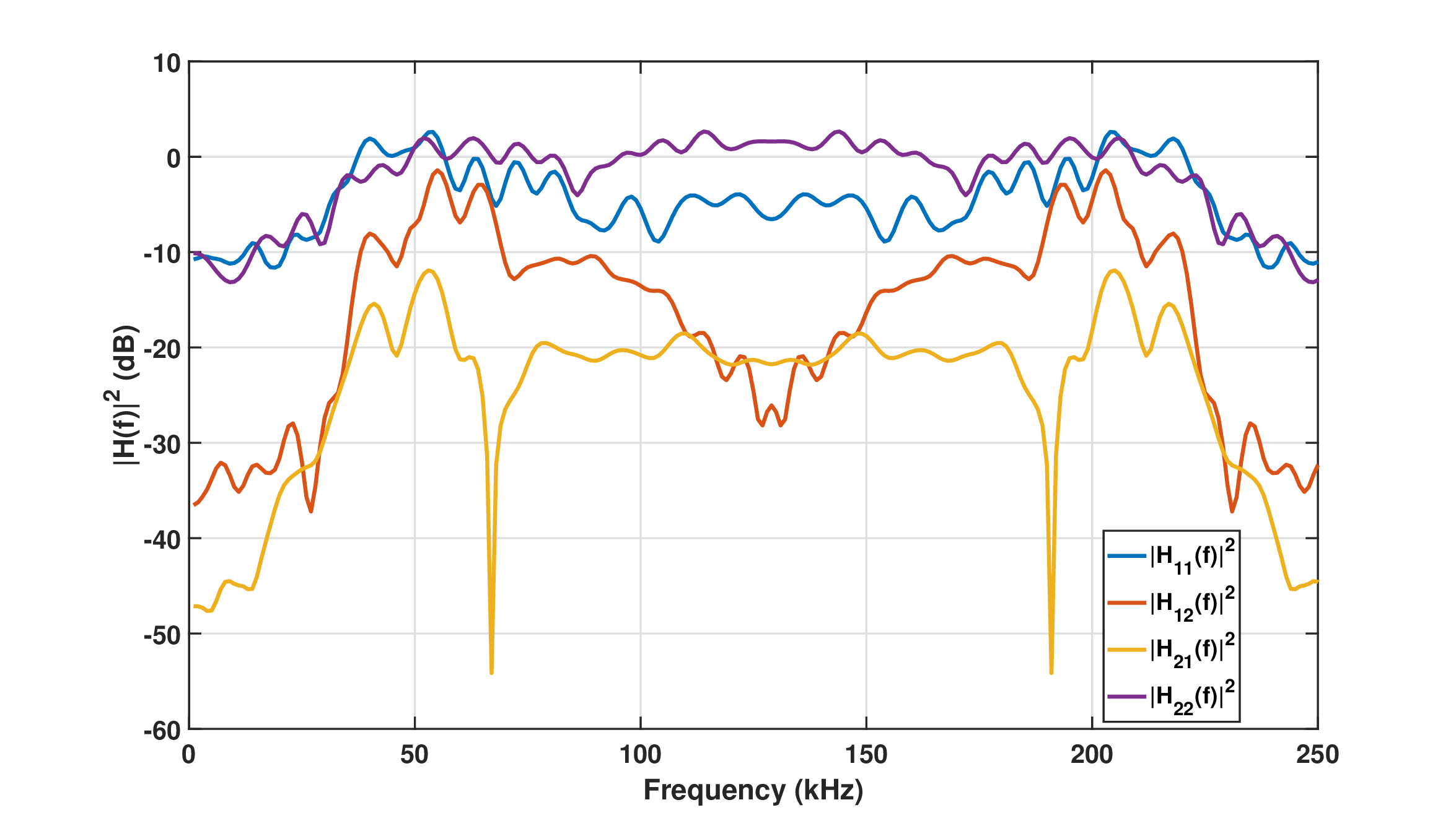}
	\caption{Measured channel frequency response for $2\times2$ MIMO-PLC in CENELEC band.}
	\label{fig:cfr-2by2-plc}
        
\end{figure}

Channel measurements considered in this work are the same as in \cite{ChannelModel}. The measurements were conducted in the laboratory over an LV power line in the frequency range $0 - 200\,$kHz with inter-carrier spacing equal to $1.5625\,$kHz. The corresponding channel frequency response (CFR) is shown in Fig.~\ref{fig:cfr-2by2-plc}.

\begin{figure}
\centering
	\includegraphics[width=1\columnwidth]{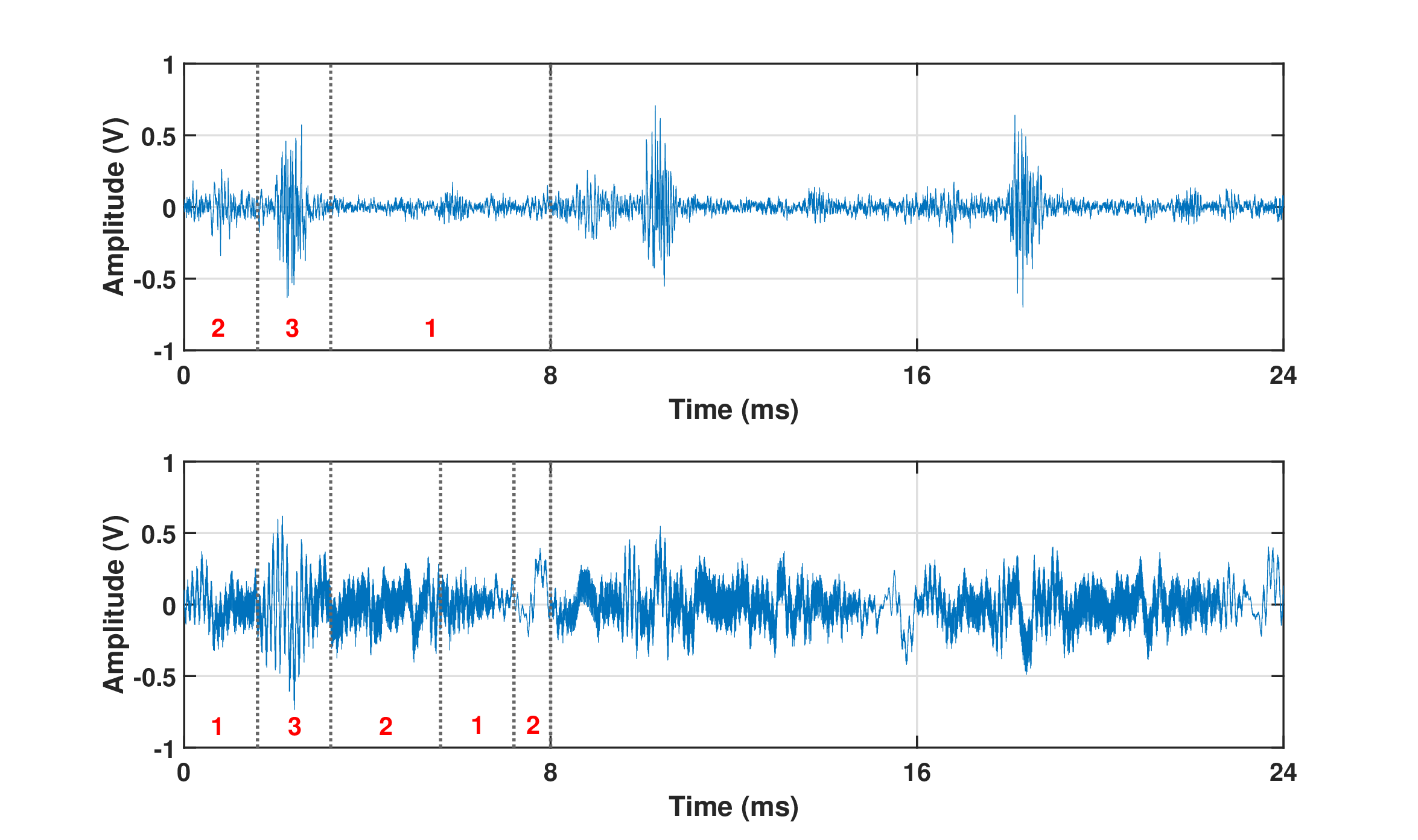}
    \caption{Three cycles of cyclostationary noise of $2\,$$\times$$\,2$ MIMO NB-PLC. Top figure indicates noise generated by FRESH filtering and the bottom figure shows the measured noise. Red numbers specify the associated class to each section.}
	\label{fig:noise-fresh-3cycle}
    
\end{figure}

The two cyclostationary noise sample sets used in this work are the noise measured in the laboratory using an oscilloscope at $1.2$ MHz sampling rate and the noise generated by the FRESH filtering model proposed in \cite{elgenedy2016frequency} and described in Sec.~\ref{subsubsec:FRESH-MIMO}. Here, the FRESH filter consists of $K\,$=$\,19$ branches to shift the input signals. Moreover, four sets of filter coefficients $ {g^{(1,1)}_{k}}[l], {g^{(1,2)}_{k}}[l], {g^{(2,1)}_{k}}[l], {g^{(2,2)}_{k}}[l]$ are considered for each branch of the $2\times2$ MIMO FRESH filter. These coefficients are generated based on real measurements of the NB-PLC noise in the laboratory. Each set of coefficients is a vector with length equal to $65$.
The following analysis is performed over a downsampled version of the noise at $400\,$kHz corresponding to the sampling rate considered in the CENELEC-A and CENELEC-B frequency bands. 
Figure~\ref{fig:noise-fresh-3cycle} depicts three noise periods of one phase of each model for a $2\,$$\times$$\,2$ MIMO system. As shown in this figure, the noise is cyclostationary with a periodicity of half the AC cycle. With a sampling frequency of $400\,$kHz and AC cycle equal to $62.5\,$Hz, we have $3,200$ samples per period. Consequently, each noise period coincides with $10$ transmitted OFDM symbols.
%

\subsection{Classification of the Noise}
As it can be seen in Fig.~\ref{fig:noise-fresh-3cycle}, the duration of each noise period is equal to $8\,$ms, there are distinct regions with specific characteristics that are repeated in every period and are called classes in \cite{moaveninejad2019gaussian}. Based on the IEEE standard 1901.2 \cite{standard}, the highest accuracy can be obtained by dividing classes into a specific number of OFDM symbols. It is worth noting that it is possible to consider a lower number of samples per slot but, in this case, there would be a higher number of classes in one period. However, according to what is suggested in IEEE standard 1901.2 \cite{standard}, two or three regions are sufficient to characterize the temporal and spectral behavior of the PLC noise. This classification is done based on changes in statistics of the samples and was introduced for the first time in \cite{moaveninejad2019gaussian}.
Each region is considered as a separate class. All the samples belonging to a specific class are collected from $N_\text{period} = 20 $ periods to analyze the behavior. 
The classification algorithm is given in Algorithm~\ref{Algorithm:noise-classification}, where the two thresholds $th_1$ and $th_2$ are chosen empirically in such a way that for all periods, the same noise slots are allocated to a certain class.
\begin{algorithm}
\caption{Cyclostationary noise classification}\label{Algorithm:noise-classification}
\begin{algorithmic}[1]
\State \textbf{inputs:} \,$\{z[n]\}, N_\text{period}, N_\text{s}, N_\text{fft}, N_\text{cp}, th_{1}, th_{2}$.
\For{\texttt{$s=1,\ldots, N_\text{period} \times N_\text{s}$}}
    \State$z^{s} = z[n]$, \hspace{.2cm} $n=(N_\text{fft}+N_\text{cp})(s-1)+1$,\ldots,$(N_\text{fft}+N_\text{cp})s$
    \State$\sigma_{z^{s}} = \sqrt{\frac{\sum_{n=1}^{(N_\text{cp}+N_\text{fft})}(z[n]- \mu_{z^{s}})^2}{{(N_\text{fft}+N_\text{cp})}}} $
    \State${\sigma}^{s} = {{\sigma}_{{z}^{s}}} \hspace{.2cm}$ 
\EndFor
\State${\sigma}_\text{min}= \min_{s\in \{1,\ldots,N_\text{period} \times N_\text{s}\}} \left \{{{\sigma}^{s}}\right \} $
\For{\texttt{$s=1,\ldots,N_\text{s}$}}
    \State${D}^{s} = {\sigma}^{s}-{\sigma}_\text{min}$ 
    \If{${D}^{s} \leq th_{1}$} 
        \State$index_\text{Class~1} \gets s$
    \ElsIf{$th_{1} < {D}^{s} \leq th_{2}$}
        \State$index_\text{Class~2} \gets s$
    \ElsIf{$th_{2} < {D}^{s}$}
        \State$index_\text{Class~3} \gets s$
    \EndIf
\EndFor
\State Record  the  index  of  noise  slots  associated  with  each class
\end{algorithmic}
\end{algorithm}

Algorithm~\ref{Algorithm:noise-classification} consists of the following steps:
\begin{itemize}
    \item Step 1: In each period, divide the sequence of noise samples $\{z[n]\}$ into $N_\text{s}$ slots, each consisting of $N_\text{fft}+N_\text{cp}$ samples.
    \item Step 2: Estimate the standard deviation ${\sigma}^{s}$ of each noise slot $s$, with $s=1,\ldots,N_\text{period}$$\,\times\,$$N_\text{s}$.
    \item Step 3: Compute ${D}^{s}$, the difference between the standard deviation of each single slot and the minimum standard deviation of all slots in $N_\text{period}$ periods.
    \item Step 4: Classify noise slots into three classes based on two thresholds $th_{1}$ and $th_{2}$.
\end{itemize}
\begin{figure*}
    \centering
    \includegraphics[width=0.8\textwidth]{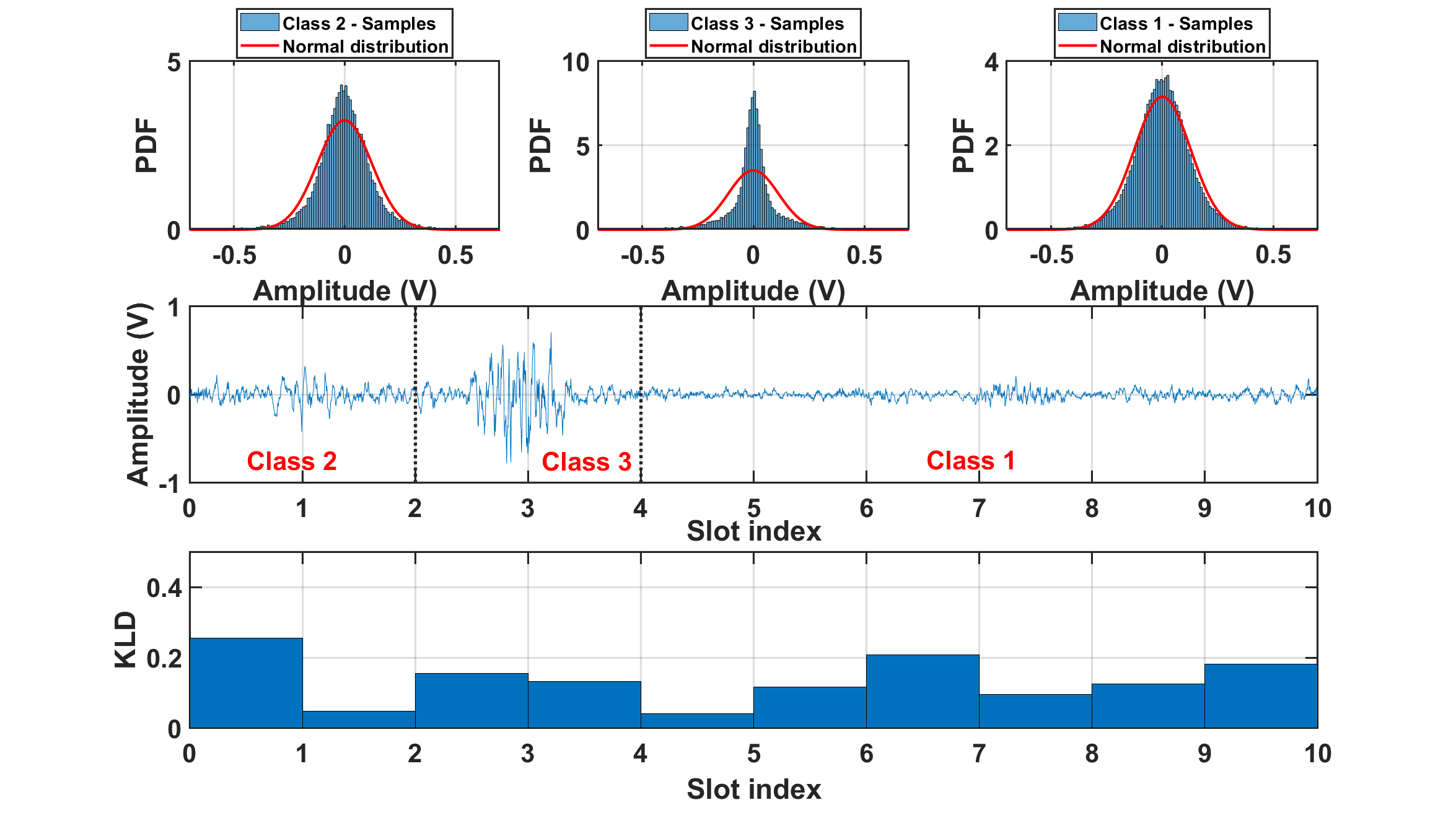}
    \caption{Analyzing the FRESH noise: pdf of each class, and KL divergence of each slot.}
    \label{fig:KLD_pdf123}
\end{figure*}

Figure~\ref{fig:KLD_pdf123} shows that the FRESH noise classification in each cycle leads to allocating the last six slots of noise to class~1, the first two slots to class~2, and the remaining slots to class~3. Hence, for $N_\text{period}$ periods, the total number of samples in each class $i$ is equal to $({N_{\text{s}}})_i \times (N_{\text{p}}+N_{\text{cp}}) \times N_\text{period}$.
After assigning each noise slot to a specific class, we collect all noise samples from $20$ periods that belong to that class. Class~2 and class~3 have $12800$ samples each and $38400$ samples belong to class~1. Then, similar to \cite{moaveninejad2019gaussian}, the histogram of each class is obtained and shown on top of Fig \ref{fig:KLD_pdf123}.

\begin{figure*}
    \centering
    \includegraphics[width=0.8\textwidth]{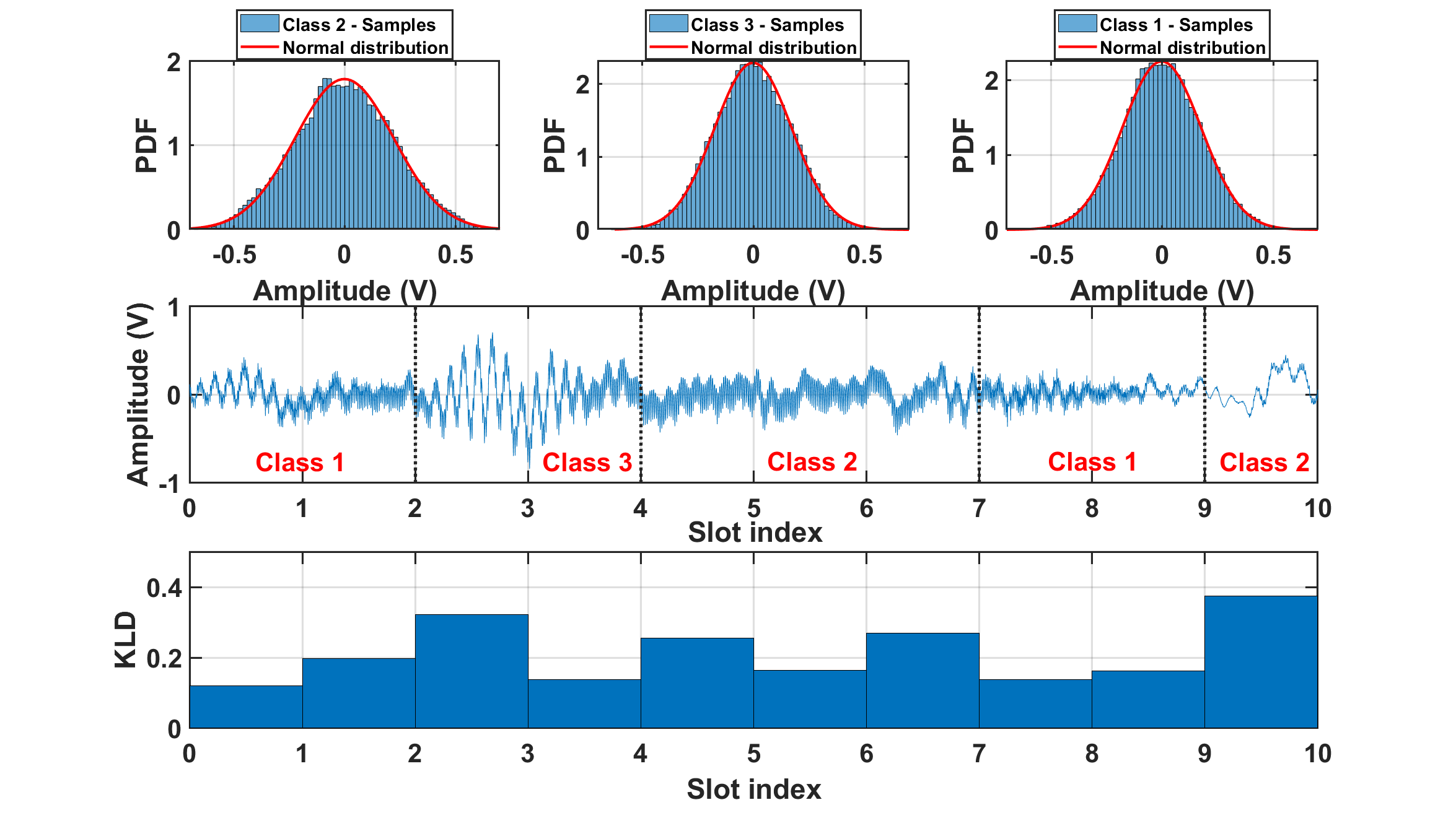}
    \caption{Analyzing the measured noise: PDF of each class, and KL divergence of each slot.}
    \label{fig:KLD_pdf123m}
\end{figure*}

Figure~\ref{fig:KLD_pdf123m} is the resulting classification of the measured noise. The process is applied directly to the noise with the original sampling frequency of $1.2\,$MHz, equal to $10^{4}$ samples per period. Accordingly, $N_{\text{s}}$ is set to $20$ slots each containing $N_{\text{p}}+N_{\text{cp}} = 500$ samples to perform finer classification according to Algorithm~\ref{Algorithm:noise-classification}. In this case, the PSD is also used to identify different frequency components which contribute to the noise to examine the effect of high and low frequency components. This method helped us to further identify the underlying contributors that influence the classification outcome.
After classifying the measured noise, we downsample it to $400\,$kHz so as to comply with the sampling frequency of the FRESH model. Accordingly, by considering $20$ periods and the distribution of the classes as in Fig.~\ref{fig:KLD_pdf123m}, $25600$ samples are allocated to class~1 and class~2 each, and $12800$ samples to class~3.

\subsection{Gaussianity Test for the Noise}
As mentioned in Sec.~\ref{sec:gaussianity_test}, we need to test the Gaussianity of the noise samples and find the length of temporal portions in each period which contains samples with normal distribution. Consequently, this allows us to use the Shannon formula for evaluating the capacity for each portion. We apply the KLD test to the whitened cyclostationary noise and the results are shown in Fig.~\ref{fig:KLD_pdf123} for the FRESH model and in Fig.~\ref{fig:KLD_pdf123m} for the measured noise.


We test the Gaussianity by selecting different values for $A$ used in calculating $N_{\text{p}}=[\frac{N_{\text{fft}}+N_{\text{cp}}}{A}]-N_{\text{cp}}$. When $A$ is set to $1$, we have $N_{\text{p}} = N_\text{fft}$ which means that the length of each Gaussian portion appended with CP is equal to the length of one slot. Here, we chose $\text{KLD}_\text{th}= 0.4$ as the threshold for Gaussianity. Since KLD of all slots are below this threshold, we keep $A=1$ but in case that for any slot KLD goes beyond this value, we increase $A$ to decrease the length of portions until all the portions respect the threshold. Here, the variance $\hat{\sigma}_{z^{p+\text{cp}}}^{2}=\hat{\sigma}_{z^{s}}^{2}$ and mean value $\hat{\mu}_{z^{p+\text{cp}}} = \hat{\mu}_{z^s}$ are estimated from $q^{p}_{z}(x)$ by collecting samples of each slot from $N_\text{period}=20$. Afterwards, we compute KLD for each portion using $N_\text{itr}=10^{3}$ iterations and then averaging the result.

\subsection{Effect of Whitening on the Gaussianity of Each Portion of Time}

\begin{figure}
\centering
    \includegraphics[width=1\columnwidth]{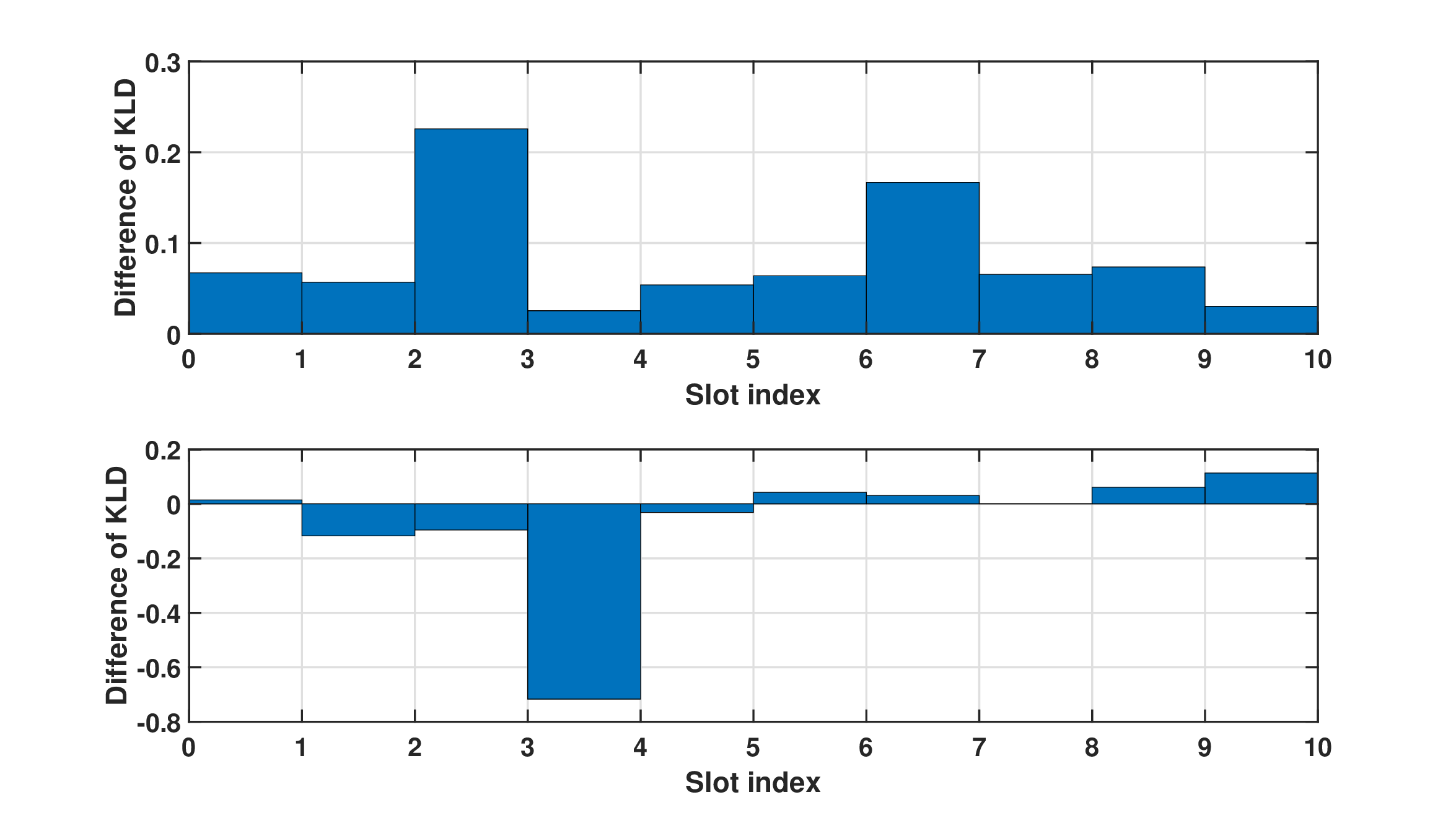}
    \caption{Amount of change in Gaussianity based on the difference between KLD before and after whitening ($\text{KLD}_{\text{before}}-\text{KLD}_{\text{after}}$) for each time portion with $N_{\text{p}}=N_{\text{fft}}=256$. Top figure indicates measured noise and the bottom figure shows the noise generated using the FRESH filtering model.}
	\label{fig:KLD_difference}
    
\end{figure}

Figure~\ref{fig:KLD_difference} shows the difference between KLD of different portions of the noise for both FRESH filtering and measured noise samples, computed before and after whitening. It is observed that whitening decreases Gaussianity in all the slots in the case of measured noise while for the FRESH filtering model, the change does not follow a specific pattern. As a result, by approximating the behavior of the noise after whitening as Gaussian and by keeping the Gaussianity threshold mentioned in the last subsection, we evaluate a lower bound to the capacity. It is worth mentioning that the closeness of the bound to real capacity depends on the distance between the actual distribution of the noise samples and a normal distribution. In the next subsection, we evaluate these capacity bounds for each portion.
 
\subsection{Lower Bound on Capacity for MIMO NB-PLC in Each Gaussian Portion of Time}
\label{subsec:Capacity}
For a $2\times2$ MIMO transmission, \eqref{eq:y_i} is written as 
\begin{align}
    y^{i}_n = \sum_{j=1}^{2}\sum_{l=0}^{64}h^{(i,j)}[l]x^{j}[n-l]+z^{i}_n,\ i \hspace{-.1cm} = \hspace{-.1cm} 1,\hspace{-.05cm}2,
\label{eq:y_i_2}
\end{align}
or in matrix notation as
\begin{align}
    \begin{bmatrix}
    y^{1}[n]\\y^{2}[n]
   \end{bmatrix}
= \sum_{l=0}^{64}
    \begin{bmatrix}
    h^{(1,1)}[l] & h^{(1,2)}[l] \\ 
    h^{(2,1)}[l] & h^{(2,2)}[l] 
   \end{bmatrix} 
   \begin{bmatrix}
   x^{1}[n-l]\\x^{2}[n-l]
   \end{bmatrix}
+   \begin{bmatrix}
   z^{1}[n]\\z^{2}[n]
   \end{bmatrix}.
    \label{eq:Yn_2}
\end{align}
For each portion $p$ consisting of $N_{\text{p}}$ samples,  \eqref{eq:Yn_2} is expanded into \eqref{eq:y_big} for linear convolution.
\begin{figure*}
	{\footnotesize
		\begin{align}
	&	\begin{bmatrix}
		\begin{bmatrix}
		y^{1}[n]\\y^{2}[n]
		\end{bmatrix}
		\\ \vdots \\
		\begin{bmatrix}
		y^{1}[n+N_{\text{p}}-1]\\y^{2}[n+N_{\text{p}}-1]
		\end{bmatrix}
		\end{bmatrix} = 
		\begin{bmatrix}
		\begin{bmatrix}
		h^{(1,1)}[l] & h^{(1,2)}[l] \\ 
		h^{(2,1)}[l] & h^{(2,2)}[l]
		\end{bmatrix}_{l=L-1}
		& H[L\hspace{-.02cm}-\hspace{-.02cm}2] & \cdots & \cdots &  H[0] & 0 &\cdots &  \cdots\\
		0 & H[L\hspace{-.02cm}-\hspace{-.02cm}1] & \cdots & \cdots & \cdots & H[0]  & 0 & \cdots\\
		\vdots & \ddots & \ddots & \ddots & &  & \ddots &\vdots\\
		0 & \cdots & 0 & H[L\hspace{-.02cm}-\hspace{-.02cm}1] & \cdots & \cdots & \cdots & H[0]
		\end{bmatrix}
		\begin{bmatrix}
		\begin{bmatrix}
		x^{1}_{[n\hspace{-.02cm}-\hspace{-.02cm}(L\hspace{-.02cm}-\hspace{-.02cm}1)]}\\x^{2}_{[n\hspace{-.02cm}-\hspace{-.02cm}(L\hspace{-.02cm}-\hspace{-.02cm}1)]}
		\end{bmatrix}
		\\ \vdots \\
		\begin{bmatrix}
		x^{1}[n]\\x^{2}[n]
		\end{bmatrix}
		\\ \vdots \\
		\begin{bmatrix}
		x^{1}_{[n+N_{\text{p}}-1]}\\x^{2}_{[n+N_{\text{p}}-1]}
		\end{bmatrix}
		\end{bmatrix}
		\notag\\
 &	+   \begin{bmatrix}
		\begin{bmatrix}
		z^{1}[n]\\z^{2}[n]
		\end{bmatrix}
		\\ \vdots 
		\label{eq:y_big}\\
		\begin{bmatrix}
		z^{1}[n+N_{\text{p}}-1]\\z^{2}[n+N_{\text{p}}-1]
		\end{bmatrix}
		\end{bmatrix}
		\end{align}}
\end{figure*}
Then, we use the $(2 \times N_{\text{p}}) \times (2 \times N_{\text{p}})$ noise autocorrelation matrix of MIMO system for noise spatio-temporal whitening. Subsequently, we evaluate the lower bound on capacity, which we refer to here as capacity for simplicity, as described in Sec.~\ref{subsec:capacity}. Capacity of each noise model has been computed both in the presence of Channel State Information at the Transmitter (CSIT) and absence of CSIT.



\begin{figure*}
\begin{subfigure}{0.48\textwidth}
	\includegraphics[width=\columnwidth]{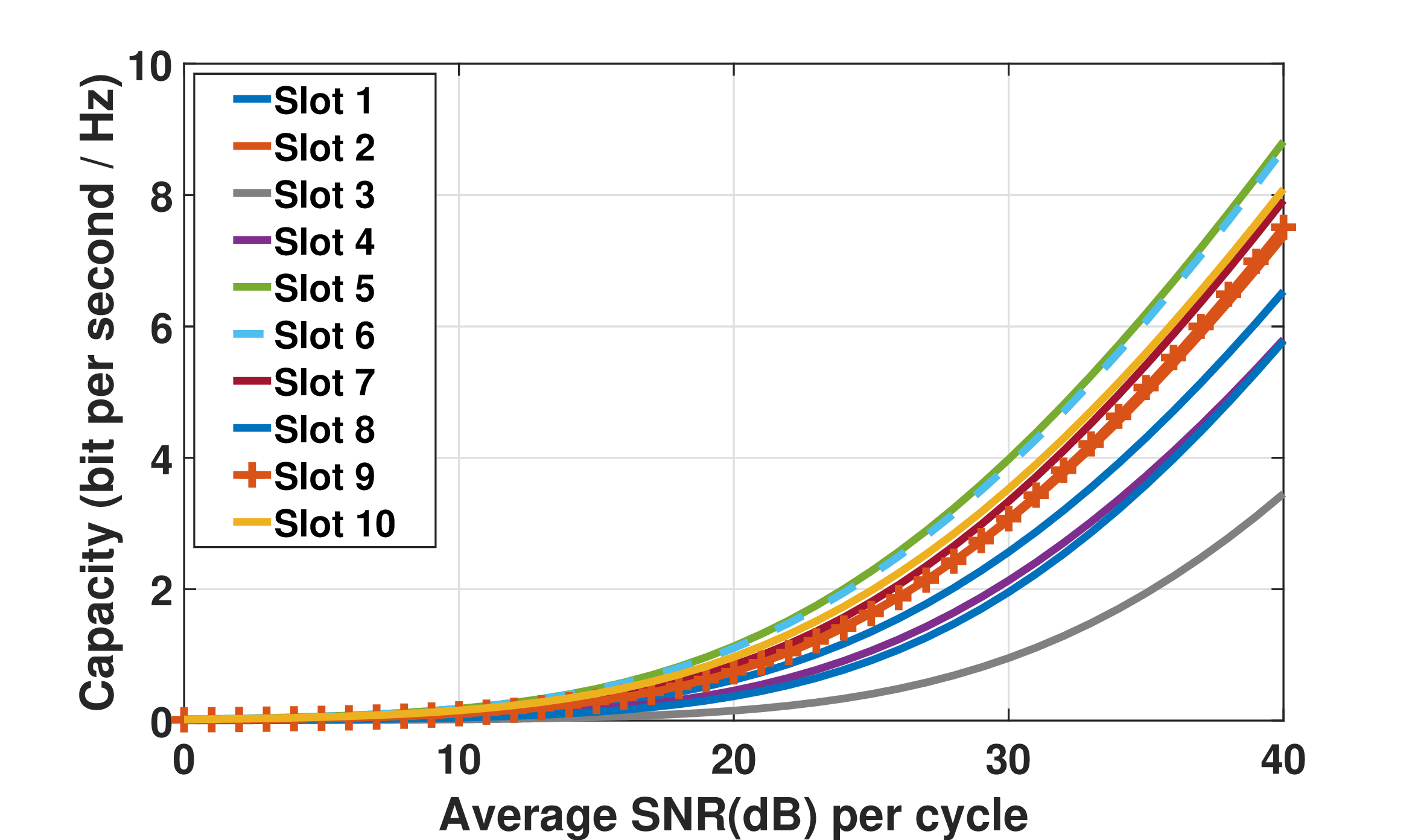}
	\caption{Capacity per slot of FRESH noise without CSIT}
	\label{fig:cap_FRESH_slot_time_unknown}
\end{subfigure}
\hfill
\begin{subfigure}{0.48\textwidth}
	\includegraphics[width=\columnwidth]{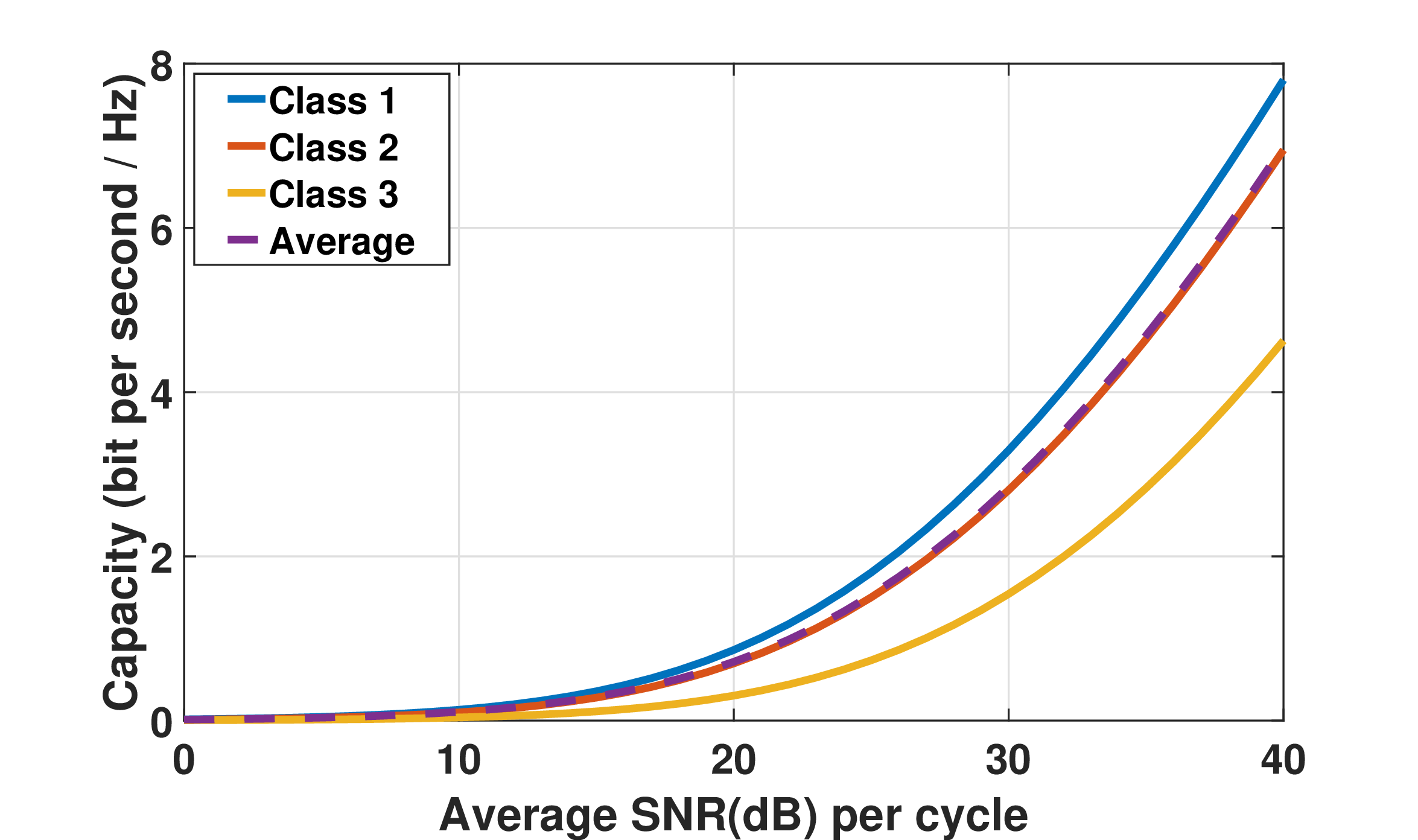}
	\caption{Capacity per class of FRESH noise without CSIT}
	\label{fig:cap_FRESH_class_time_unknown}
\end{subfigure}
\vskip\baselineskip
\vspace{-0.5cm}
\begin{subfigure}{0.48\textwidth}
	\includegraphics[width=\columnwidth]{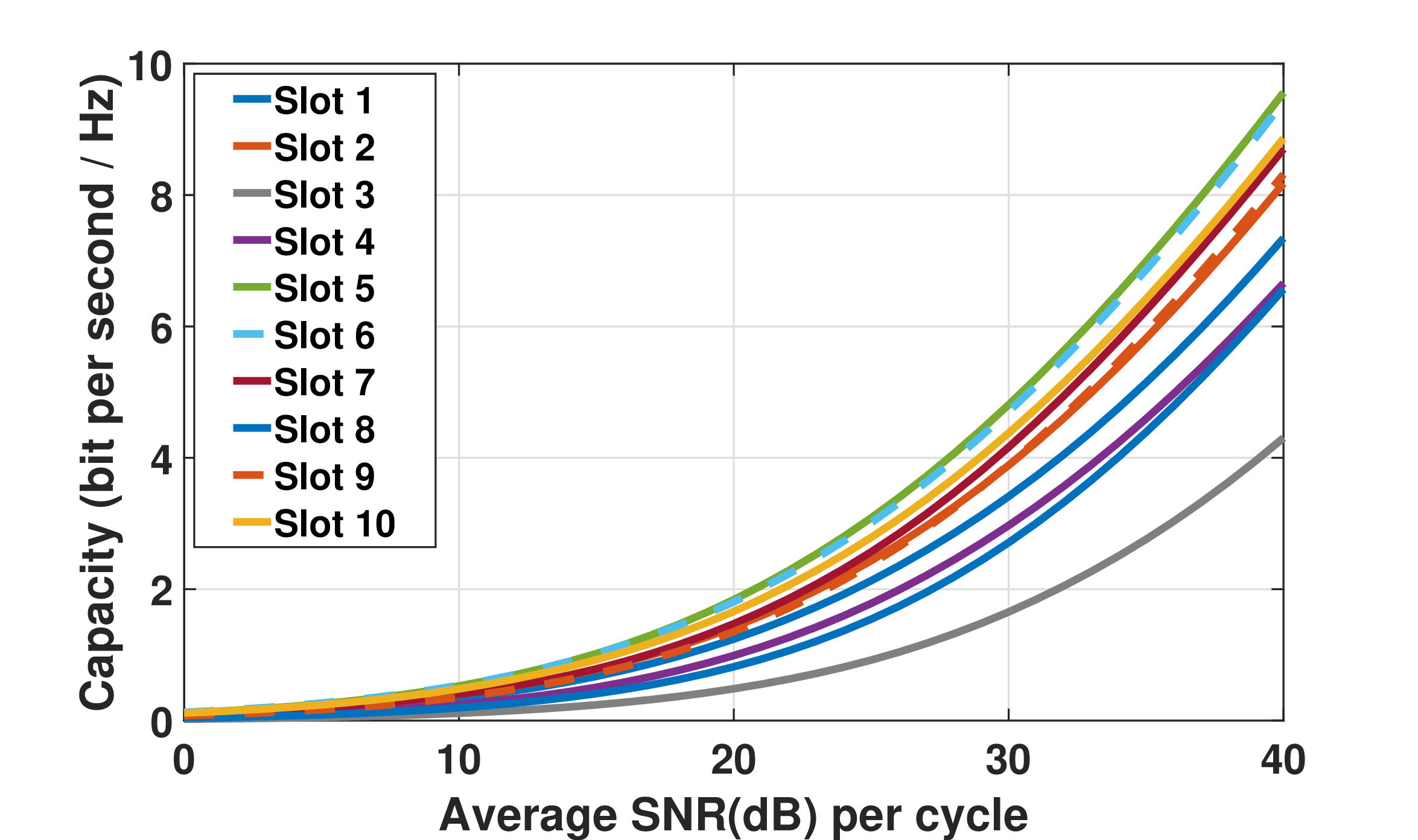}
	\caption{Capacity per slot of FRESH noise with CSIT}
	\label{fig:cap_FRESH_slot_time_unknown_CSIT}
\end{subfigure}
\hfill
\begin{subfigure}{0.48\textwidth}
	\includegraphics[width=\columnwidth]{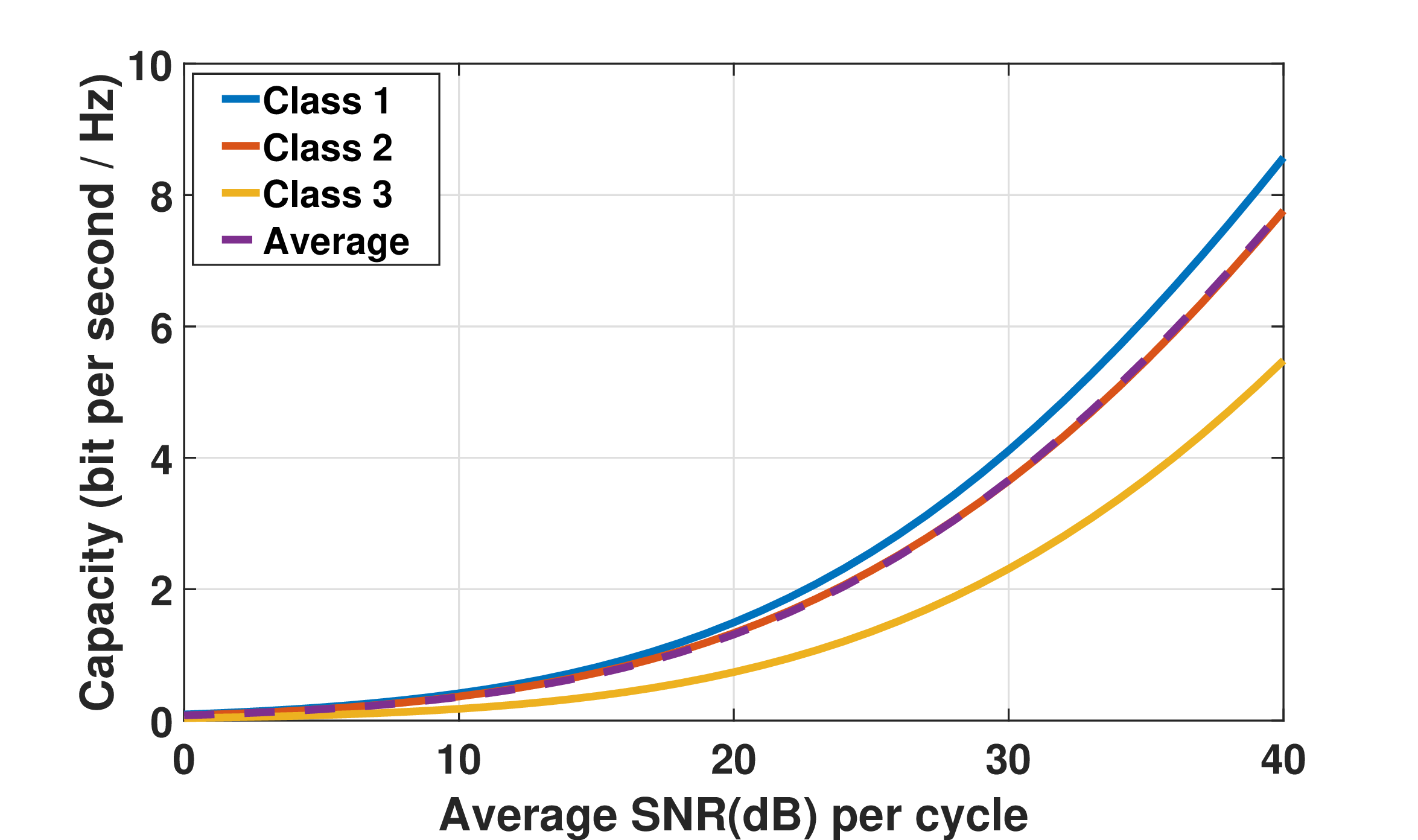}
	\caption{Capacity per class of FRESH noise with CSIT}
	\label{fig:cap_FRESH_class_time_unknown_CSIT}
\end{subfigure}
\vskip\baselineskip
\vspace{-0.5cm}
\begin{subfigure}{0.48\textwidth}
	\includegraphics[width=\columnwidth]{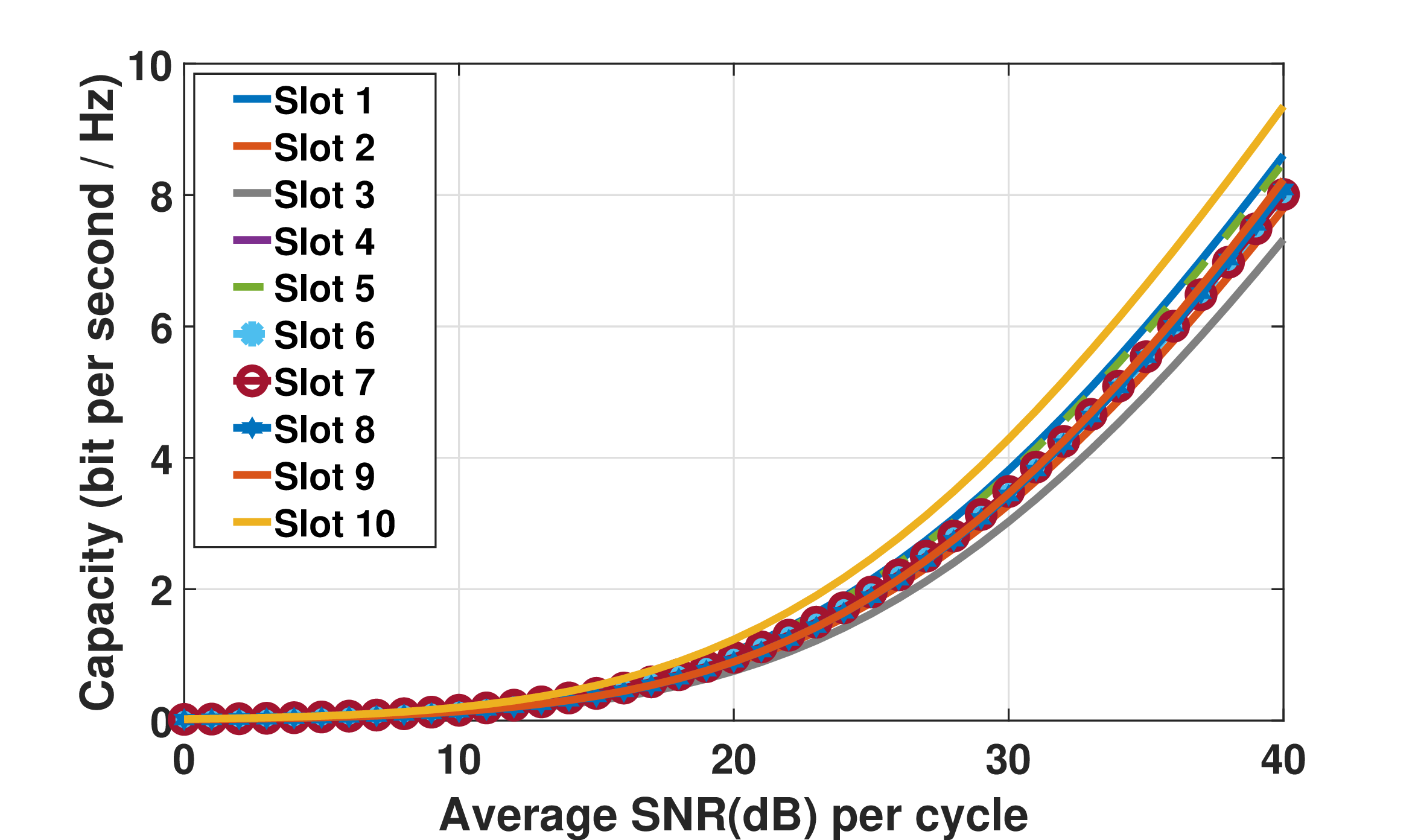}
	\caption{Capacity per slot of measured noise without CSIT}
	\label{fig:cap_measured_slot_time_unknown}
\end{subfigure}
\hfill
\begin{subfigure}{0.48\textwidth}
	\includegraphics[width=\columnwidth]{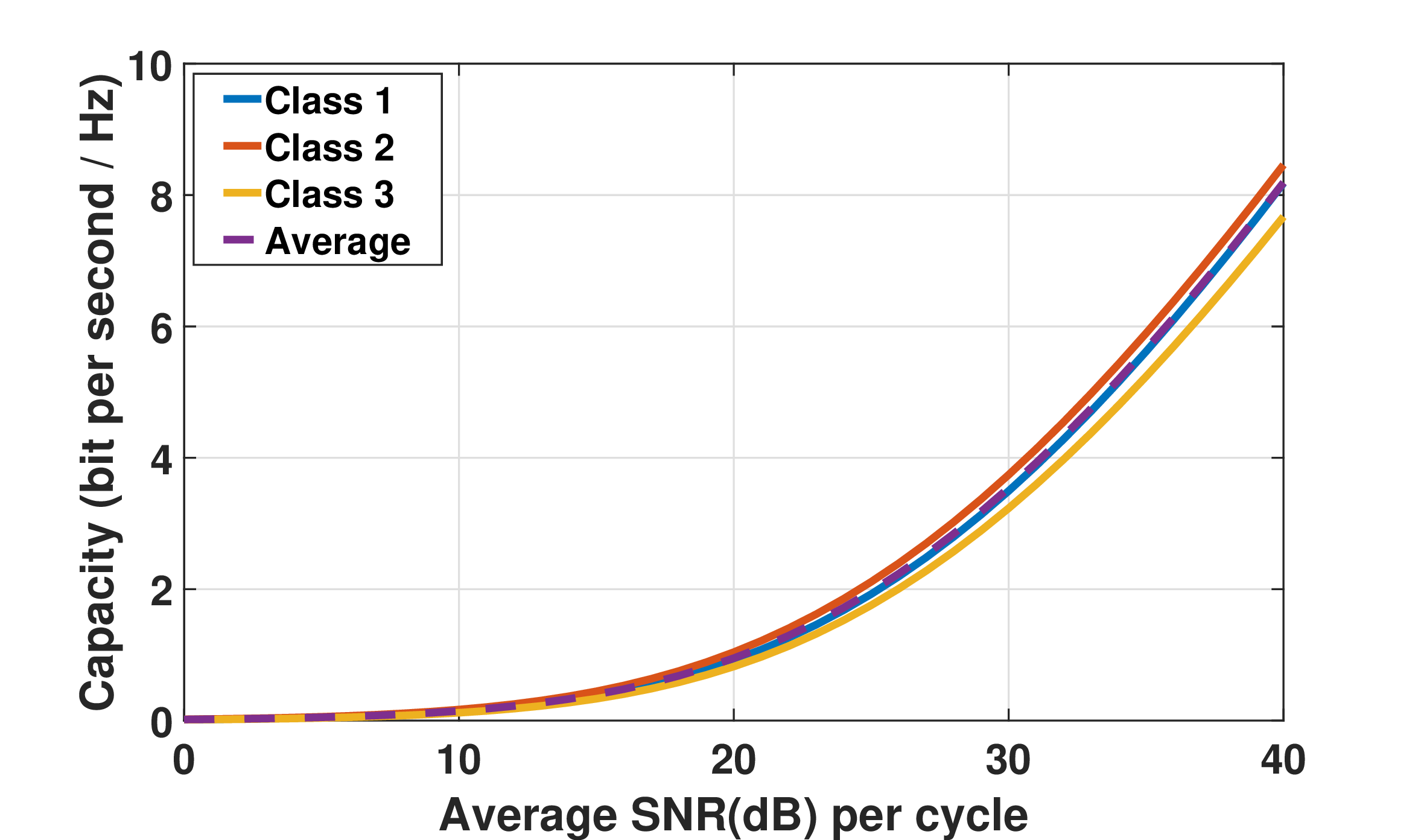}
	\caption{Capacity per class of measured noise without CSIT}
	\label{fig:cap_measured_class_time_unknown}
\end{subfigure}	
\vskip\baselineskip
\vspace{-0.5cm}
\begin{subfigure}{0.48\textwidth}
	\includegraphics[width=\columnwidth]{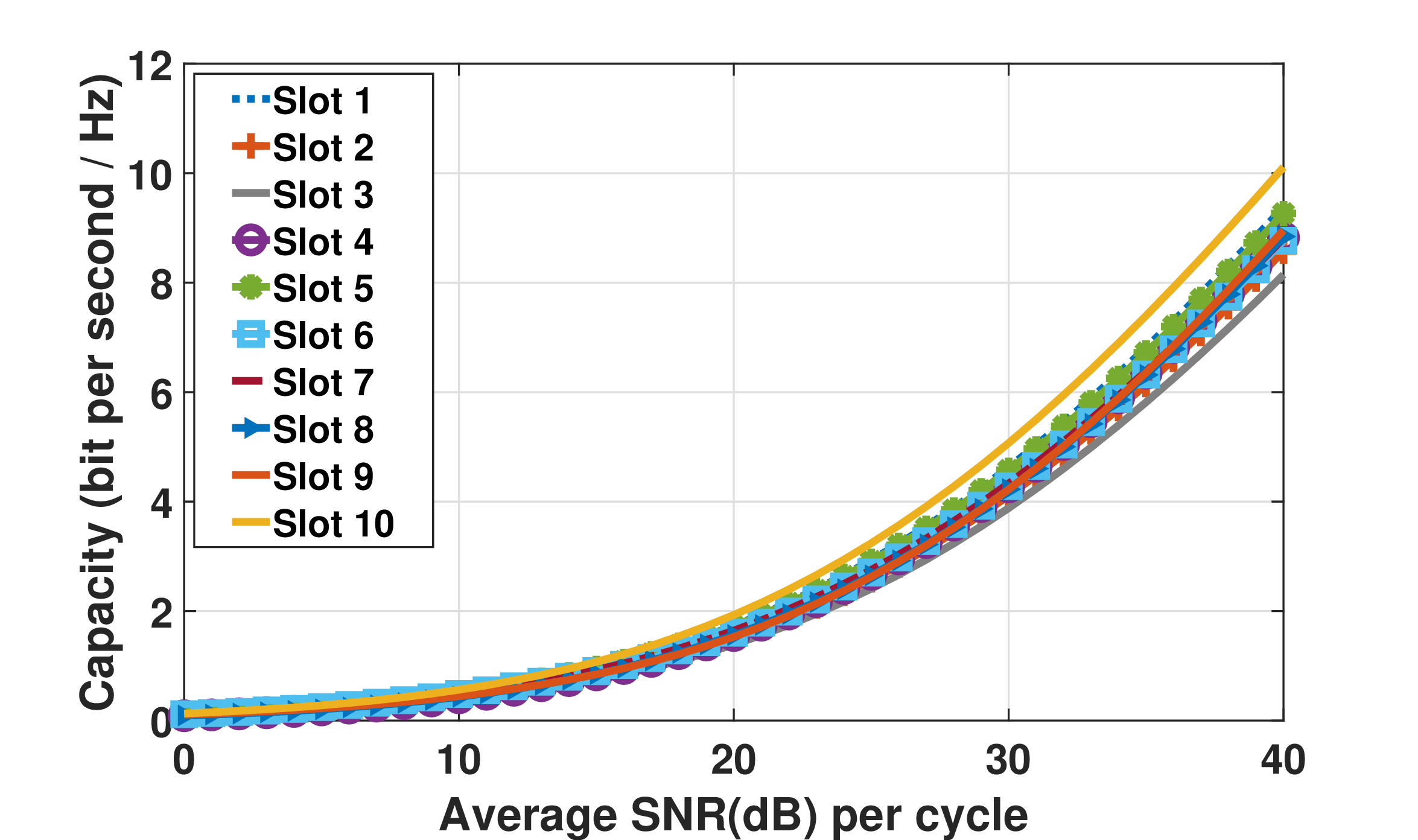}
	\caption{Capacity per slot of measured noise with CSIT}
	\label{fig:cap_measured_slot_time_unknown_CSIT}
\end{subfigure}	
\hfill
\begin{subfigure}{0.48\textwidth}
	\includegraphics[width=\columnwidth]{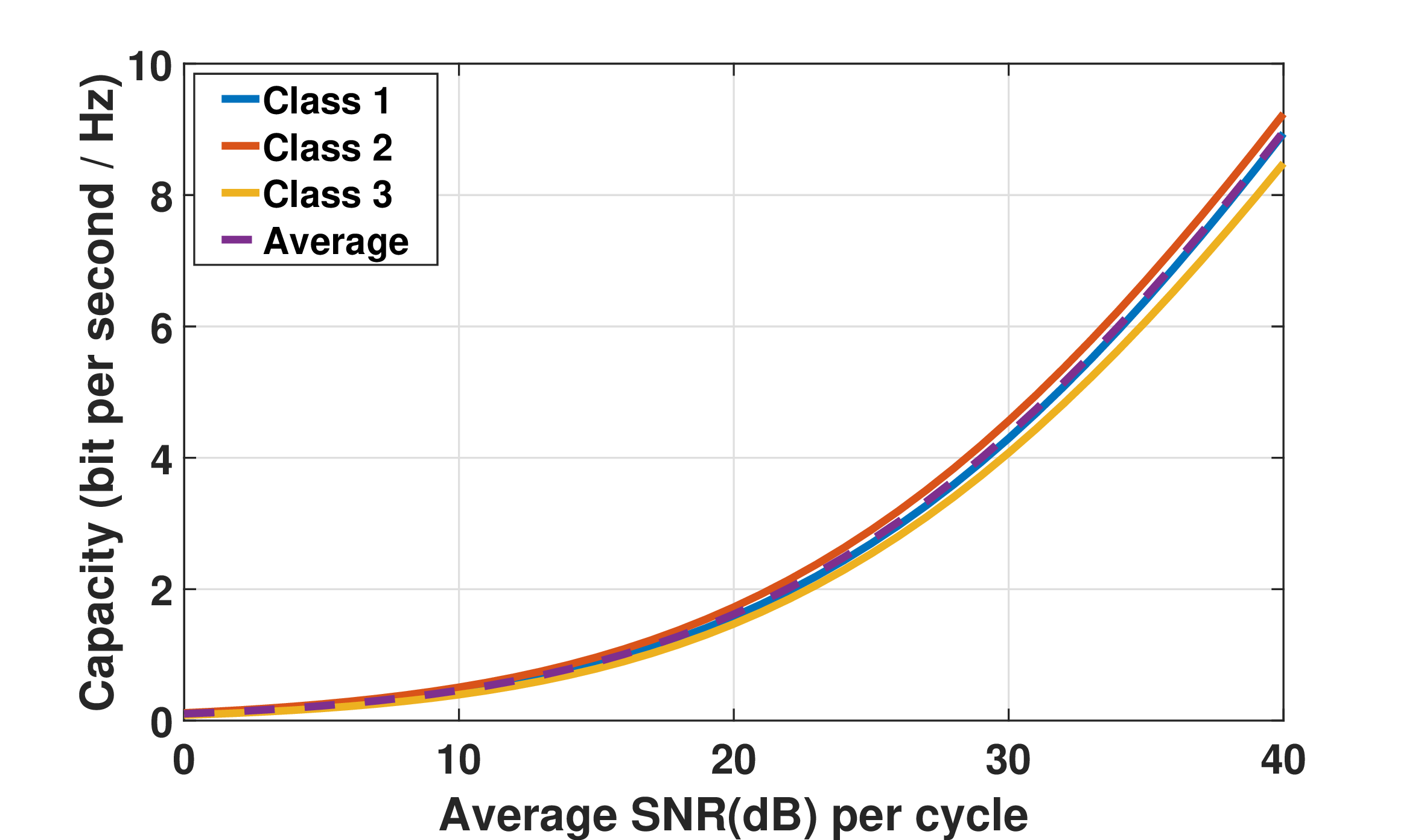}
	\caption{Capacity per class of measured noise with CSIT}
	\label{fig:cap_measured_class_time_unknown_CSIT}
\end{subfigure}
\caption{Capacity lower bounds of the measured and FRESH noise data sets in presence of CSIT and in absence of CSIT.}
\end{figure*}

From Fig.~\ref{fig:cap_FRESH_slot_time_unknown} to Fig.~\ref{fig:cap_measured_class_time_unknown_CSIT} we can see that for an average SNR at the receiver in the presence of cyclostationary noise, there are different capacities corresponding to different slots since each slot experiences a different level of noise. Consequently, for each average received SNR, there are three different values of SNR for data samples belonging to each class.

Figure~\ref{fig:cap_FRESH_slot_time_unknown} and Fig.~\ref{fig:cap_measured_slot_time_unknown} show the capacity in absence of CSIT for each slot for the FRESH filtering and measured noise, respectively. In this case, the power is equally divided between $10$ slots in each period. We can observe that the slots with more impulsive distribution of the noise have lower capacity. Due to the fact that measured noise samples demonstrate a more Gaussian behavior in almost all portions, the capacity curves of these slots are closer to each other in comparison with the FRESH filtering noise. In addition, the lowest capacity in both cases is related to slot~3 due to the spikiness of its noise distribution. Another aspect to consider is the slope with which the capacity curves increase with the increase of SNR. This slope is almost the same for all the slots of the measured noise and the curves start to distance themselves from the low capacity values at about $10\,$dB. While in the FRESH filtering model, the slope is different for different slots and the detachment from low capacity values happens at different values of SNR between $10$ to $20\,$dB.

In the presence of CSIT, which enables optimum power allocation, the capacity is reported in Fig.~\ref{fig:cap_FRESH_slot_time_unknown_CSIT} and Fig.~\ref{fig:cap_measured_slot_time_unknown_CSIT} for the FRESH filtering and measured noise samples, respectively. The optimum power allocation increases the capacity, especially at lower SNR values which can be noted from the distance of capacity curves from the SNR axis. In this case, the slope of the capacity curve is steeper and the convexity of the curves has decreased, which means that there is also some improvement for higher values of SNR. We note that the compactness of the capacity curves is almost identical as in the case of no CSIT, which means that for measured noise samples the curves are in a bundle, while for the FRESH filtering model they are distant from each other. Despite the improvement in the capacity of slot~3, it is still the lowest among all the other slots in both noise sample sets.

Since each class of the noise consists of a certain number of slots, capacity of each class is obtained by the average capacity of the slots belonging to that class as shown in Figs.~\ref{fig:cap_FRESH_class_time_unknown} and~\ref{fig:cap_measured_class_time_unknown} for no CSIT, and Figs.~\ref{fig:cap_FRESH_class_time_unknown_CSIT} and~\ref{fig:cap_measured_class_time_unknown_CSIT} in the case of known CSIT for FRESH filtering and measured noise samples, respectively. The behavior of impulsive distribution of the noise in slots contributing to class~3 can be noted from its lower capacity with respect to the other two classes. We can also see the vicinity of the capacity curves of the classes to each other and to their average in measured noise samples since all the classes demonstrate an almost normal distribution but this is not the case for the FRESH filtering model. In the latter, the curves are far from each other especially class~3 which has the lowest capacity. The same observations can be made for waterfilling case as for the capacity of the slots. Here, we can easily see the earlier detachment of the curves from the low capacity values and also the fact that the capacity has improved even for high values of SNR.

\section{Conclusion And Future Work}
\label{sec:Conclusion}
In this paper, the problem of deriving the capacity of MIMO NB-PLC system is addressed when the noise is correlated in the time and spatial domain. We used laboratory measured and MIMO FRESH filter generated noise with such characteristics to numerically evaluate a lower bound on the capacity. Moreover, the noise does not show unique behavior in each period. After classification of the cyclostationary noise into three classes, the corresponding distributions show Gaussian, moderate impulsive, and strong impulsive behaviors for the FRESH generated noise and Gaussian for the measured noise. First, we divide each period into temporal portions with Gaussian behavior and then perform spatio-temporal whitening on each portion and evaluate the capacity. The length of each portion is set to the length of each OFDM symbol after Gaussianity test. However, the length of each portion could be divided to fractions of an OFDM symbol until all portions in one period satisfy the Gaussianity test.

Simulation results show the difference between measured noise and the one generated by FRESH filtering. Authors in \cite{elgenedy2016frequency} stated that the accuracy of the FRESH model could be improved by increasing the length of the FIR filter and the number of branches. However, this enhancement introduces heightened complexity in noise generation. A solution to this issue is presented in \cite{simpler_FRESH}. This approach, besides spectral shaping, considers a temporal shaping that allows for better control over the amount of impulsiveness of the noise classes, thereby enhancing its versatility. Authors in \cite{simpler_FRESH} developed their model based on the FRESH filtering noise samples, however an area for future investigation could be the improvement of the model by utilizing the actual measured noise samples for fitting. 

There is also a difference between the capacity obtained using these two models and Gaussian noise. This difference is greater in class~3, which exhibits strong impulsive behavior. By decreasing the size of the OFDM symbols, it is possible to fine-tune the boundaries between each class. We observed that, in this way, the capacity of the FRESH model approaches the capacity of the measured data. Furthermore, investigation of the simulation results show that the overall average capacity is doubled in the MIMO case compared to SISO, while the improvement for class~3 is more than double. Our analysis will help to design the optimal receiver in terms of bit loading using waterfilling algorithm. This is the main reason to extend the current work by applying it for adaptive modulation. Modulation order could be adapted to the capacity evaluated for each slot or each class of noise due to different levels of average SNR at the receiver of NB-PLC.

\bibliographystyle{ieeetr}
\bibliography{biblio/Capacity.bib}
  
\end{document}